\documentclass[10pt]{article}
\usepackage[numbers]{natbib}
\usepackage{amsmath}
\usepackage{amsfonts}
\usepackage{amsopn}
\usepackage{epsfig}
\usepackage{ifthen}
\usepackage[frozencache=true,cachedir=.]{minted}
\usepackage{url}
\usepackage{multirow}
\usepackage{siunitx}

\usepackage{graphicx}
\graphicspath{{.}{../icml2018/diagrams/}}

\newif\ifsvd
\svdtrue  

\newcommand{\field}[1]{\mathbb{#1}}

\newcommand{\R}{\field{R}}

\newcommand{\vect}[1]{\boldsymbol{#1}} 
\newcommand{\mat}[1]{\boldsymbol{#1}} 
\newcommand{\tvect}[1]{\tilde{\boldsymbol{#1}}}
\newcommand{\tmat}[1]{\tilde{\boldsymbol{#1}}}
\newcommand{\tscal}[1]{\tilde{#1}}
\newcommand{\hvect}[1]{\hat{\boldsymbol{#1}}}
\newcommand{\hmat}[1]{\hat{\boldsymbol{#1}}}
\newcommand{\hscal}[1]{\hat{#1}}
\newcommand{\bvect}[1]{\bar{\boldsymbol{#1}}}
\newcommand{\bmat}[1]{\bar{\boldsymbol{#1}}}
\newcommand{\bscal}[1]{\bar{#1}}
\newcommand{\vzero}{\vect{0}}

\newcommand{\dummystring}{QWERTYU}
\newcommand{\vci}[3][\dummystr]{\ifthenelse{\equal{#1}{\dummystring}}{\vect{#2}_{#3}}{\vect{#2}_{#3}^{(#1)}}}
\newcommand{\mx}[3][\dummystr]{\ifthenelse{\equal{#1}{\dummystring}}{\mat{#2}_{#3}}{\mat{#2}_{#3}^{(#1)}}}
\newcommand{\tvci}[3][\dummystr]{\ifthenelse{\equal{#1}{\dummystring}}{\tvect{#2}_{#3}}{\tvect{#2}_{#3}^{(#1)}}}
\newcommand{\tmx}[3][\dummystr]{\ifthenelse{\equal{#1}{\dummystring}}{\tmat{#2}_{#3}}{\tmat{#2}_{#3}^{(#1)}}}
\newcommand{\tscl}[3][\dummystr]{\ifthenelse{\equal{#1}{\dummystring}}{\tscal{#2}_{#3}}{\tscal{#2}_{#3}^{(#1)}}}
\newcommand{\hvci}[3][\dummystr]{\ifthenelse{\equal{#1}{\dummystring}}{\hvect{#2}_{#3}}{\hvect{#2}_{#3}^{(#1)}}}
\newcommand{\hmx}[3][\dummystr]{\ifthenelse{\equal{#1}{\dummystring}}{\hmat{#2}_{#3}}{\hmat{#2}_{#3}^{(#1)}}}
\newcommand{\hscl}[3][\dummystr]{\ifthenelse{\equal{#1}{\dummystring}}{\hscal{#2}_{#3}}{\hscal{#2}_{#3}^{(#1)}}}
\newcommand{\bvci}[3][\dummystr]{\ifthenelse{\equal{#1}{\dummystring}}{\bvect{#2}_{#3}}{\bvect{#2}_{#3}^{(#1)}}}
\newcommand{\bmx}[3][\dummystr]{\ifthenelse{\equal{#1}{\dummystring}}{\bmat{#2}_{#3}}{\bmat{#2}_{#3}^{(#1)}}}
\newcommand{\bscl}[3][\dummystr]{\ifthenelse{\equal{#1}{\dummystring}}{\bscal{#2}_{#3}}{\bscal{#2}_{#3}^{(#1)}}}

\DeclareMathOperator{\trace}{tr}
\DeclareMathOperator{\diag}{diag}
\DeclareMathOperator*{\argmax}{arg max}

\DeclareMathOperator*{\sgn}{sgn}

\DeclareMathOperator{\sym}{sym}
\DeclareMathOperator{\rank}{rk}

\DeclareMathOperator{\tril}{tril}
\DeclareMathOperator{\triu}{triu}

\newcommand{\Ind}[1]{\mathrm{I}_{\{#1\}}}

\newcommand{\Id}{\mat{I}}

\newcommand{\eps}{\mathrm{\varepsilon}}

\newcommand{\figref}[1]{Figure~\ref{fig:#1}}
\newcommand{\tabref}[1]{Table~\ref{tab:#1}}
\newcommand{\secref}[1]{Section~\ref{sec:#1}}
\renewcommand{\eqref}[1]{Eq.~\ref{eq:#1}}
\newcommand{\eqp}[1]{(\ref{eq:#1})}

\newcommand{\tsm}[2][\dummystring]{\tscl[#1]{m}{#2}}

\newcommand{\va}[2][\dummystring]{\vci[#1]{a}{#2}}
\newcommand{\vb}[2][\dummystring]{\vci[#1]{b}{#2}}
\newcommand{\vc}[2][\dummystring]{\vci[#1]{c}{#2}}
\newcommand{\vd}[2][\dummystring]{\vci[#1]{d}{#2}}

\newcommand{\vf}[2][\dummystring]{\vci[#1]{f}{#2}}

\newcommand{\vh}[2][\dummystring]{\vci[#1]{h}{#2}}

\newcommand{\vu}[2][\dummystring]{\vci[#1]{u}{#2}}
\newcommand{\vv}[2][\dummystring]{\vci[#1]{v}{#2}}
\newcommand{\vw}[2][\dummystring]{\vci[#1]{w}{#2}}
\newcommand{\vx}[2][\dummystring]{\vci[#1]{x}{#2}}
\newcommand{\vy}[2][\dummystring]{\vci[#1]{y}{#2}}
\newcommand{\vz}[2][\dummystring]{\vci[#1]{z}{#2}}

\newcommand{\vth}[2][\dummystring]{\vci[#1]{\theta}{#2}}
\newcommand{\vmu}[2][\dummystring]{\vci[#1]{\mu}{#2}}

\newcommand{\vlam}[2][\dummystring]{\vci[#1]{\lambda}{#2}}

\newcommand{\vsigma}[2][\dummystring]{\vci[#1]{\sigma}{#2}}

\newcommand{\bva}[2][\dummystring]{\bvci[#1]{a}{#2}}
\newcommand{\bvb}[2][\dummystring]{\bvci[#1]{b}{#2}}
\newcommand{\bvc}[2][\dummystring]{\bvci[#1]{c}{#2}}
\newcommand{\bvd}[2][\dummystring]{\bvci[#1]{d}{#2}}

\newcommand{\bvlam}[2][\dummystring]{\bvci[#1]{\lambda}{#2}}

\newcommand{\mxa}[2][\dummystring]{\mx[#1]{A}{#2}}
\newcommand{\mxb}[2][\dummystring]{\mx[#1]{B}{#2}}
\newcommand{\mxc}[2][\dummystring]{\mx[#1]{C}{#2}}

\newcommand{\mxe}[2][\dummystring]{\mx[#1]{E}{#2}}
\newcommand{\mxf}[2][\dummystring]{\mx[#1]{F}{#2}}
\newcommand{\mxg}[2][\dummystring]{\mx[#1]{G}{#2}}

\newcommand{\mxk}[2][\dummystring]{\mx[#1]{K}{#2}}
\newcommand{\mxl}[2][\dummystring]{\mx[#1]{L}{#2}}
\newcommand{\mxm}[2][\dummystring]{\mx[#1]{M}{#2}}

\newcommand{\mxq}[2][\dummystring]{\mx[#1]{Q}{#2}}
\newcommand{\mxr}[2][\dummystring]{\mx[#1]{R}{#2}}
\newcommand{\mxs}[2][\dummystring]{\mx[#1]{S}{#2}}

\newcommand{\mxu}[2][\dummystring]{\mx[#1]{U}{#2}}
\newcommand{\mxv}[2][\dummystring]{\mx[#1]{V}{#2}}

\newcommand{\mxx}[2][\dummystring]{\mx[#1]{X}{#2}}
\newcommand{\mxy}[2][\dummystring]{\mx[#1]{Y}{#2}}

\newcommand{\mxsigma}[2][\dummystring]{\mx[#1]{\Sigma}{#2}}

\newcommand{\mxlam}[2][\dummystring]{\mx[#1]{\Lambda}{#2}}

\newcommand{\tmxm}[2][\dummystring]{\tmx[#1]{M}{#2}}

\newcommand{\hmxk}[2][\dummystring]{\hmx[#1]{K}{#2}}

\newcommand{\bmxa}[2][\dummystring]{\bmx[#1]{A}{#2}}
\newcommand{\bmxb}[2][\dummystring]{\bmx[#1]{B}{#2}}
\newcommand{\bmxc}[2][\dummystring]{\bmx[#1]{C}{#2}}

\newcommand{\bmxl}[2][\dummystring]{\bmx[#1]{L}{#2}}

\newcommand{\bmxq}[2][\dummystring]{\bmx[#1]{Q}{#2}}
\newcommand{\bmxr}[2][\dummystring]{\bmx[#1]{R}{#2}}

\newcommand{\bmxu}[2][\dummystring]{\bmx[#1]{U}{#2}}
\newcommand{\bmxv}[2][\dummystring]{\bmx[#1]{V}{#2}}

\newcommand{\bmxlam}[2][\dummystring]{\bmx[#1]{\Lambda}{#2}}

\newcommand{\rng}[2][1]{{#1},\dots,{#2}}

\title{Auto-Differentiating Linear Algebra}
\author{Matthias Seeger, Asmus Hetzel, Zhenwen Dai, Eric Meissner, \\
        Neil D. Lawrence \\
        {\em matthis@amazon.de}, {\em ahhetzel@amazon.de}, {\em zhenwend@amazon.co.uk}, \\
        {erimeiss@amazon.co.uk}, {\em lawrennd@amazon.co.uk}}

\begin{document}

\maketitle\thispagestyle{empty} 

\begin{abstract}
Development systems for {\em deep learning} (DL), such as Theano, Torch, TensorFlow, or MXNet, are easy-to-use tools for creating complex neural network models. Since gradient computations are automatically baked in, and execution is mapped to high performance hardware, these models can be trained end-to-end on large amounts of data. However, it is currently not easy to implement many basic machine learning primitives in these systems (such as Gaussian processes, least squares estimation, principal components analysis, Kalman smoothing), mainly because they lack efficient support of linear algebra primitives as differentiable operators. We detail how a number of matrix decompositions (Cholesky, LQ, symmetric eigen) can be  implemented as differentiable operators. We have implemented these primitives in MXNet, running on CPU and GPU in single and double precision. We sketch use cases of these new operators, learning Gaussian process and Bayesian linear regression models, where we demonstrate very substantial reductions in implementation complexity and running time compared to previous codes. Our MXNet extension allows end-to-end learning of hybrid models, which combine {\em deep neural networks} (DNNs) with Bayesian concepts, with applications in advanced Gaussian process models, scalable Bayesian optimization, and Bayesian active learning.
\end{abstract}

\section{Introduction}
\label{sec:intro}

Deep neural networks, trained on vast amounts of data, have recently revolutionized several machine learning applications, ranging from object classification and detection in computer vision over large vocabulary speech recognition to machine translation of natural language text. Apart from large amounts of data and high-performance computing hardware, a major driver of this success has been the flexibility and ease of use of modern {\em deep learning development systems} (DLDS), such as Theano, Torch, TensorFlow, or MXNet. These systems map algorithms to {\em computation graphs}, where nodes are multi-dimensional arrays, and vertices are differentiable operators. Given this abstraction, gradients w.r.t.\ any node are obtained {\em automatically} by way of reverse mode differentiation, a generalization of error backpropagation. While some DLDS maximize flexibility by unfolding the graph on the fly, others may apply graph optimization in order to minimize runtime or memory usage on a specific target hardware. All a user has to do is to specify this symbolic graph, or even just re-combine existing components, and bind it to input data. Not only has this simple means of specifying a complex model lowered the technical bar to entry into cutting-edge ML, it is also dramatically shortening the time from idea to large data experimentation, which is so crucial for data-driven innovation.

Given this major advance in ease of modeling, it is remarkable to note how many ``pre-deep-learning'' algorithmic primitives remain out of reach of current DLDS. Here are some examples, which can be found in any ML standard textbook \cite{Bishop:06, Barber:12, Murphy:12, Rasmussen:06}:
\begin{itemize}
\item
  Least squares estimation, solving linear systems
\item
  Gaussian process models
\item
  Principal components analysis, linear discriminant analysis,
  canonical correlation analysis
\item
  Kalman filtering and smoothing in linear dynamical systems
\end{itemize}
Why are none of these easily available in most DLDS, as are LSTM or Inception modules? Inspecting the former algorithms, they make use of certain linear algebra primitives, such as Cholesky decomposition, backsubstitution, LQ decomposition, symmetric eigendecomposition, or singular value decomposition. As we will demonstrate in this work, {\em a limited number of linear algebra primitives is all that is missing} in order to cover a substantial number of ML cornerstone algorithms, {\em as long as they are available as differentiable operators}.

In this work, we show how the following linear algebra primitives can be incorporated as operators into a DLDS:
\begin{itemize}
\item
  Cholesky decomposition, backsubstitution
\item
  LQ decomposition (transpose of QR decomposition)
\item
  Symmetric eigendecomposition
\ifsvd
\item
  Singular value decomposition (SVD)
\fi
\end{itemize}
Much like matrix-matrix multiplication or 2D convolution in current DLDS, these new operators are treated as {\em first-class citizens}, in that (a) they support automatic differentiation like any other node in a computation graph, and (b) map to highly tuned implementations for CPU and GPU architectures commonly used by ML practitioners.

We contributed implementations for all operators detailed here to the MXNet deep learning development system, where they are found in {\tt mxnet.sym.linalg} and {\tt mxnet.nd.linalg}.\footnote{
  \url{https://github.com/apache/incubator-mxnet}}
While currently less popular than TensorFlow or PyTorch, MXNet offers some advantages for our purposes. As an Apache open source project, MXNet invites contributions outside the current DNN mainstream. Via the novel Gluon API, users can work both with static (as in TensorFlow) and dynamic graphs (as in PyTorch). MXNet lacked linear algebra primitives, allowing us to start from a clean slate. Finally, MXNet is highly optimized for reduced runtime and memory footprint, its low-level operator API allows to make best use of in-place capabilities offered by underlying linear algebra libraries. In contrast, both TensorFlow and PyTorch already have levels of abstraction in place (see \secref{rel-software}).

Our implementations work both on CPUs and GPUs, and can be used in {\tt float64} or {\tt float32}. It is built on standard linear algebra APIs (BLAS, LAPACK), for which highly tuned implementations are available for all common CPUs and NVidia GPUs. By exploiting in-place capabilities in BLAS, we avoid overhead due to allocation of temporary memory. In a range of empirical comparisons, our MXNet operators are more than three times faster on CPU (Intel MKL build), and consistently faster on GPU, than equivalents in TensorFlow. Our work has already been used to scale up Bayesian optimization and meta-learning \cite{Perrone:17}.

The structure of this paper is as follows. We summarize related work and outline our goals in \secref{relwork}. In \secref{examples}, we sketch several machine learning applications enabled by our novel operators: Gaussian process regression, Bayesian linear regression, and Kalman filtering. Complete example notebooks are available for download, as referenced there. Forward and backward expressions for a range of advanced linear algebra operators are detailed in \secref{operators}, and our implementation of these operators in MXNet is described in \secref{mxnet}.  In \secref{exper}, we present a range of empirical evaluations, comparing the runtime of our operators to equivalents in TensorFlow, as well as benchmarking our implementation on advanced sparse GP and variational auto-encoded deep GP models. Full derivations of all backward expressions are provided in the Appendix.

\section{Related Work. Goals}
\label{sec:relwork}

In this section, we review prior work on reverse mode differentiation for matrix decompositions and differentiable operators in DLDS. Note that what we denote {\em backward} in the remainder of this work, is called {\em pullback} in the automatic differentiation community.

A number of different algorithms for the Cholesky decomposition are proposed in \cite{Murray:16}. The backward expression we use here, appears in this work. The author recommends to implement a block version of the operator, for reasons of cache-friendly memory access. However, no code or empirical results are provided. A number of backward expressions for linear algebra are given in \cite{Giles:08}, among them the one we use for the symmetric eigendecomposition here. The author does not provide code. Finally, a number of results for forward and reverse mode differentiation of matrix decompositions are provided in \cite{Walter:12}. This work contains pullbacks for the symmetric eigendecomposition (same as ours) and the QR decomposition. For the latter, they concentrate on the case $\mxa{} = \mxq{}\mxr{}\in\R^{m\times n}$, where $m\ge n$, $\mxq{}\in\R^{m\times m}$, and $\mxr{}\in\R^{m\times n}$ is upper triangular and padded with zero rows. In typical machine learning applications, we have $m\gg n$, and an $m\times m$ matrix could not be stored. Their expression is different to ours. They do not provide code. In summary, while most of the backward expressions used in our work here are not novel (a possible exception is our expression for the LQ decomposition), they are not widely known in the ML community. Moreover, none of the work above provides serious implementations on top of standard libraries like BLAS and LAPACK.

\subsection{Software}
\label{sec:rel-software}

Here, we summarize the current implementations of linear algebra operators in commonly used DLDS. We concentrate on operators which can be first-class citizens in a computation graph, because gradients w.r.t.\ inputs are supported.

We start with {\em TensorFlow} \cite{Abadi:15}. The operators relevant to the current work are {\tt tf.cholesky} (our {\tt potrf}), {\tt tf.matrix\_triangular\_solve} (our {\tt trsm}), {\tt tf.qr} (related to our {\tt gelqf}), and {\tt tf.self\_adjoint\_eig} (our {\tt syevd}). Code for backward expressions is at \url{https://github.com/tensorflow/tensorflow/blob/master/tensorflow/python/ops/linalg_grad.py}:
\begin{itemize}
\item
  {\tt \_CholeskyGrad} (our {\tt potrf\_backward}): Implements same expression
  as ours, but not in-place (three temporary matrices are used).
\item
  {\tt \_MatrixTriangularSolveGrad} (our {\tt trsm\_backward}): Implements same
  expression as ours.
\item
  {\tt \_SelfAdjointEigV2Grad} (our {\tt syevd\_backward}): Implements same
  expression as ours, but not in-place (at least three temporary matrices are
  used).
\item
  {\tt \_QrGrad} (related to our {\tt geqlf\_backward}): Implements an expression
  which is more complex than ours. Not in-place, at least five temporary matrices
  are used.
\ifsvd
\item
  {\tt \_SvdGrad}: Backward expression for the singular value decomposition (SVD).
  Supports the operator {\tt tf.svd}.
\fi
\end{itemize}
At a higher level, linear algebra in TensorFlow is mapped to {\em Eigen} (\url{eigen.tuxfamily.org}). While this simplifies implementations, it is wasteful in terms of runtime and memory. The TensorFlow implementation is not optimized for in-place computations, and a substantial number of temporary matrices are used. It is unclear whether this is due to the dependence on Eigen, or could be fixed.

Results of runtime comparisons between TensorFlow and our MXNet implementation are given in \secref{exper-tf}, for a number of operators, both on CPU and GPU. Our MXNet implementation is at least three times faster on CPU (Intel MKL build), and is consistently faster on GPU as well.

Another commonly used DLDS is {\em Theano} \cite{Bergstra:10}. Here, the relevant operators are {\tt slinalg.Cholesky} (our {\tt potrf}), {\tt slinalg.solve\_lower\_triangular} (our {\tt trsm}), {\tt nlinalg.Eigh} (our {\tt syevd}). The code is at \url{http://deeplearning.net/software/theano/library/tensor/slinalg.html} and \url{http://deeplearning.net/software/theano/library/tensor/nlinalg.html}:
\begin{itemize}
\item
  {\tt slinalg.Cholesky.grad} (our {\tt potrf\_backward}): Implements the expression
  from \cite{Murray:16}, which is related to ours. Not in-place (at least three
  temporary matrices). Also, the Cholesky factor is recomputed in the backward
  pass.
\item
  {\tt slinalg.Solve.grad} (our {\tt trsm\_backward}): The argument is
  {\tt lower\_triangular}. The implementation is not in-place. Also, the
  backsubstitution is recomputed in the backward pass.
\item
  {\tt nlinalg.Eigh.grad} (our {\tt syevd\_backward}): Channeled through
  {\tt nlinalg.EighGrad}, which seems to implement an expression related to ours.
  The computation is certainly not in-place, but employs a for loop. Also, the
  eigendecomposition is recomputed in the backward pass.
\end{itemize}
While there is {\tt nlinalg.QRFull} for the QR decomposition, its gradient is not implemented. At a higher level, Theano is calling linear algebra primitives from {\em NumPy} and {\em SciPy}. These may or may not support GPU computations. The Theano implementation is not optimized for in-place computations, a substantial number of temporary matrices are used. This fact is quite likely due to the dependence on NumPy and SciPy. A further inefficiency of the Theano backward implementations is that the output of the forward computation is recomputed instead of just being passed. This does not happen in TensorFlow or MXNet.

Finally, there is {\em PyTorch} ({\tt pytorch.org}). The only relevant operator seems to be  {\tt Potrf} (our {\tt potrf}), whose code is at \url{https://github.com/pytorch/pytorch/blob/master/torch/autograd/_functions/linalg.py}. It implements the expression from \cite{Murray:16}. The code is quite wasteful, but could be improved (for example, calls to {\tt torch.gesv} could be replaced by {\tt torch.trtrs}). Note that while many BLAS and some LAPACK functions are wrapped, most do not have gradients implemented. In summary, too few linear algebra operators are implemented as first-class citizens in order to allow for use cases we are interested in there (see \secref{examples}).

The {\em GPFlow} project \cite{Matthews:17} (code at \url{https://github.com/GPflow/GPflow}) allows the user to build Gaussian process models on top of {\em TensorFlow} \cite{Abadi:15}. It is using the Cholesky decomposition and backsubstitution operators in TensorFlow, as just discussed. We demonstrate how our operators in MXNet are used to learn Gaussian process models in Sections~\ref{sec:ex-gp} and \ref{sec:ex-spgp}. Moreover, we benchmark a number of advanced models in \secref{exper}, combining sparse GP and DNN concepts in a common computation graph. Since typical GP applications operate with large dense matrices, it is crucial to avoid unnecessary temporary copies, and a careful in-place implementation such as ours should be advantageous.

In summary, while some differentiable linear algebra operators are implemented in the most popular DLDS, the support is patchy at best, with the exception of TensorFlow, which provides a broad coverage surpassing our contribution at present. In general, operators are implemented at a high level, reducing to Eigen (TensorFlow) or NumPy/SciPy (Theano), which may not take full advantage of the underlying execution context (GPU, CPU). Also, not much care is taken to avoid temporary copies. In contrast, we implement our operators into the MXNet core, calling tuned BLAS/LAPACK libraries directly. Most of our operators are implemented {\em in-place}, so that no additional memory is required. While this is somewhat more difficult to do (see \secref{mxnet}), it should pay off in applications such as those in \secref{examples}, where rather large matrices are operated on.

\subsection{Goals}

Our work of extending MXNet with differentiable linear algebra operators is motivated by the following broader goals.

\begin{itemize}
\item
  {\em Transfer key advantages of DLDS to other ML or scientific computation
  systems}.
  The main advantages of DLDS are support of high-performance compute
  platforms (GPUs, multi-threaded CPUs) and automated gradients by reverse
  mode differentiation. Frequently used non-DL machine learning packages for
  Gaussian processes or approximate Bayesian inference lack these advantages.
  Current major efforts of hardware and software developers to speed up DL can
  be leveraged for non-DL systems, if missing capabilities are added to DLDS as
  differentiable operators.
\item
  {\em High memory efficiency}.
  ML applications such as Gaussian processes or approximate Bayesian inference
  are often memory-bound, and good existing codes contain careful memory
  management. If integrated into DLDS, these memory-saving tricks have to be
  retained. Our {\tt linalg} implementation carefully leverages in-place support of
  underlying BLAS and LAPACK routines, improving on previous work.
\item
  {\em Enable end-to-end learning of hybrids between neural networks and
  Bayesian machine learning}.
  While deep neural networks are easy to construct, can be trained very efficiently,
  and learn more useful feature representations from large datasets than previous
  efforts, Bayesian methodology (such as Gaussian processes) requires less labeled
  data and works better for automated decision making (e.g., active learning,
  Bayesian optimization). While combinations can improve on either
  \cite{Wilson:16}, previous work requires implementation efforts out of scope of
  most users.\footnote{
    Bayesian concepts can be introduced to neural networks via variational
    approximations and Monte-Carlo sampling \cite{Kingma:14, Rezende:14}.
    Such methods are easily
    implemented in existing DLDS. Note that Monte-Carlo sampling is not
    competitive in conventional Bayesian ML (e.g., Gaussian processes,
    approximate inference).}
  Examples in \secref{examples} and
  \secref{exper} demonstrate that hybrid implementations can be concise and
  simple, if only a few operators are added to a DLDS.
  \citet{Perrone:17} use our work to scale up Bayesian optimization, using a
  hybrid setup with neural networks and Bayesian inference.
\end{itemize}

\section{Machine Learning Examples}
\label{sec:examples}

In this section, we provide some examples for how a small number of linear algebra operators can enable a range of machine learning methodologies inside a DLDS (we make use of MXNet). We focus on code snippets here, complete examples are provided as Jupyter notebooks referenced below.

\subsection{Gaussian Processes}
\label{sec:ex-gp}

Powerful non-parametric regression and classification models are obtained by representing unknown functions by Gaussian processes (GPs). Details about GPs can be found in \cite{Rasmussen:06}.
Here, we focus on a simple GP regression setup. Suppose we are interested in learning a function $f(\vx{})$ from data $\{ (\vx{i}, y_i)\; |\; i=\rng{n}\}$, where $y_i\in\R$ is a noisy observation of $f(\vx{i})$. A GP regression model is obtained by (a) assuming that the function $f(\vx{})$ is {\em a priori} distributed as a Gaussian process with zero mean function and covariance (or {\em kernel}) function $K(\vx{},\vx{}')$, and (b) that $y_i\sim N(f(\vx{i}),\lambda_y)$, namely targets are observed with independent Gaussian noise of variance $\lambda_y$. The kernel function $K$ has free hyper-parameters $\vth{}$. Given this model, there are two basic problems to be solved. First, we need to {\em learn} the hyper-parameters $\vth{}$ and $\lambda_y$. Second, we need to {\em predict} mean and variance of $f(\vx{*})$ at test input points $\vx{*}$. Here, we focus on the harder learning problem. It can conveniently be solved by maximizing the marginal likelihood $P(\vy{} | \vth{},\lambda_y)$ of the observed data $\vy{} = [y_1,\dots,y_n]^T$, where the unknown function $f(\cdot)$ and the noise are integrated out. We show how to express the negative log marginal likelihood $\phi$ as a symbolic MXNet expression, using the novel operators detailed in \secref{operators}. It is easy to see that
\[
  P(\vy{} | \vth{},\lambda_y) = N(\vy{} | \vzero, \mxa{}),\quad \mxa{} = \mxk{} +
  \lambda_y\Id,
\]
where $\mxk{} = [K(\vx{i},\vx{j})]_{i,j}\in\R^{n\times n}$ is the kernel matrix, and $\Id\in\R^{n\times n}$ denotes the identity matrix. As derived in \cite{Rasmussen:06}:
\[
  \phi = -\log N(\vy{}|\vzero,\mxa{}) = \frac{1}2\left( \vy{}^T\mxa{}^{-1}\vy{} +
  \log|2\pi\mxa{}| \right).
\]
Here, $|\mxx{}|$ denotes the determinant of an invertible matrix $\mxx{}$.
At this point, we use the Cholesky decomposition (\secref{oper-chol}) of $\mxa{}$:
\[
  \mxa{} = \mxl{}\mxl{}^T.
\]
Following \cite{Rasmussen:06}:
\[
  \phi = \frac{1}2\left( \|\vz{}\|^2 + n\log(2\pi) \right) + \log|\mxl{}|,\quad
  \vz{} = \mxl{}^{-1}\vy{}.
\]
Here is Python code to create an MXNet symbolic expression for $\phi$:

\begin{minted}[frame=lines]{python}
import numpy as np
from mxnet.sym.linalg import potrf, trsm
from mxnet.sym import sum, square, log, reshape

def gpregr_criterion_symbol(kern_mat, targets, noise_var, num_data,
        dtype=np.float32):
    amat = kern_mat + mat_eye_symbol(num_data, dtype) * noise_var
    lmat = potrf(amat)
    zvec = trsm(lmat, reshape(targets, shape=(-1, 1)))
    sqnorm_z = sum(square(zvec))
    logdet_l = sum(log(extract_diag_symbol(lmat, num_data)))
    return 0.5 * (sqnorm_z + (num_data*np.log(2*np.pi))) \
        + logdet_l
\end{minted}
The relevant operators are {\tt linalg.potrf} (compute Cholesky factor $\mxl{}$ of matrix $\mxa{}$) and {\tt linalg.trsm} (compute $\vz{} = \mxl{}^{-1}\vy{}$). Details for these operators are given in \secref{oper-chol}. Since $\mxl{}$ is a triangular matrix, its log determinant is
\[
  \log|\mxl{}| = \log\prod_i l_{i i} = \sum_i\log l_{i i}.
\]
Finally, {\tt extract\_diag\_symbol(lmat, num\_data)} extracts the diagonal of $\mxl{}$, and an identity matrix of size {\tt num\_data} is created by {\tt mat\_eye\_symbol(num\_data, dtype)}.

Note that the Cholesky factor $\mxl{}$ is computed once, then used both to compute $\vz{}$ and $\log|\mxl{}|$. In fact, {\tt lmat} represents a single node in the computation graph. Not only does this save memory and compute time, it is also numerically safer than computing the inverse $\mxa{}^{-1}$ and log determinant $\log|\mxa{}|$. The complete example code also demonstrates how kernel matrices are computed with {\tt linalg.syrk} and {\tt linalg.gemm2}.

Our example shows how hyper-parameter learning in a GP model can be driven by MXNet. While this is more runtime efficient and easier to implement than using existing GP software packages (see \secref{exper}), the real benefit of inserting GP operators into MXNet lies in the ability to train hybrid models end-to-end. For example, the latent function $f(\vx{})$ could be the sum of a GP and a DNN, or the kernel function $K(\vx{},\vx{}')$ could be parameterized in terms of neural networks.

A Jupyter notebook that implements the full GP model based on the above symbolic expression is available at \url{http://github.com/ARCambridge/MXNet_linalg_examples}. In the notebook, we implement the Gaussian (or Radial Basis Function, RBF) kernel, as well as the negative log likelihood criterion for GP regression, as detailed above. We also show how to compute predictive distributions for test points, after hyperparameters have been learned. The notebook is using the {\em GPy} framework (\url{https://github.com/SheffieldML/GPy}) for optimization and plotting, while all computations are done by MXNet executors. In the notebook, we demonstrate hyperparameter learning and prediction on a toy 1D dataset.

\subsection{Sparse Gaussian Processes}
\label{sec:ex-spgp}

Standard GP models are limited to medium-size datasets, since inference and learning scales cubically in the number $n$ of datapoints. One much researched direction of scaling GP models to big data are {\em sparse Gaussian process} approximations, which scale as $\mathcal{O}(n U^2)$, where $U < n$ is the number of so-called inducing points.\footnote{
  Inducing points $\vu{j}$ live in the same space as input points $\vx{i}$.}
We focus on the variational sparse GP approximation of \citet{Titsias:09}, which is principled and has led to much follow-up work. As shown there, the following criterion is an upper bound to the negative log marginal likelihood in \secref{ex-gp}:
\[
  \phi_{\text{SGP}} = -\log \mathcal{N}(\vy{} | \vzero, \hmxk{} + \lambda_y\Id) +
  \frac{1}{2\lambda_y} \trace(\mxk{f f} - \hmxk{}),
\]
where $\hmxk{} = \mxk{u f}^T\mxk{u u}^{-1}\mxk{u f}$, and $\mxk{f f}$,
$\mxk{u u}$, $\mxk{u f}$ are blocks of the joint kernel matrix over input points $\{\vx{i}\}$ and inducing points $\{\vu{j}\}$. In practice, we minimize $\phi_{\text{SGP}}$ w.r.t.\ {\em both} kernel parameters $\vth{}$, noise variance $\lambda_y$, and the inducing points (which are variational parameters). As detailed in \cite{Titsias:09}, this criterion can be computed in terms of $\mxk{u u} = \mxl{u}\mxl{u}^T$ (Cholesky factor), then $\mxb{} = \mxl{u}^{-1}\mxk{u f}$, and finally the Cholesky factor of $\mxa{} = \Id_U + \lambda_y^{-1}\mxb{}\mxb{}^T$. Given our {\tt linalg} operators, this learning criterion can be implemented very succinctly:

\begin{minted}[frame=lines]{python}
import numpy as np
from mxnet.sym.linalg import potrf, trsm, syrk, gemm2
from mxnet.sym import log, sum, square

def sgp_criterion_symbol(y, Kuu, Kuf, Kff_diag, noise_var,
        num_data, num_ind, dtype=np.float32):
    Lu = potrf(Kuu)
    B = trsm(Lu, Kuf)
    A = syrk(B) / noise_var + mat_eye_symbol(num_ind, dtype)
    La = potrf(A)
    laInvBy = trsm(La, gemm2(B, y))
    logdet_la = sum(log(extract_diag_symbol(La, num_ind)))
    return (np.log(2*np.pi)+log(noise_var))*num_data/2 \
        + logdet_la + sum(square(y))/(2*noise_var) \
        - sum(square(laInvBy))/(2*noise_var*noise_var) \
        - (sum(square(B)) - sum(Kff_diag))/(2*noise_var)
\end{minted}

Apart from {\tt linalg.potrf} and {\tt linalg.trsm}, we also use {\tt linalg.syrk} to compute $\mxb{}\mapsto \mxb{}\mxb{}^T$. This implementation constitutes a very substantial reduction and simplification of code, when compared to the reference implementation in GPy. The latter consists of hundreds of lines of code, most of which is concerned with gradient computations. These gradient expressions are typically more complex than the criterion itself, and have to be worked out by hand. Using our handful of {\tt linalg} operators, gradients are obtained automatically, removing implementation complexity and likely sources of bugs. Moreover, our experimental results in \secref{exper-sgp} indicate that even running on CPUs, sparse GP criteria and their gradients are computed much more efficiently in MXNet than in GPy.

\subsection{Least Squares Estimation. Bayesian Linear Regression}
\label{sec:ex-lse}

The linear regression problem, also known as least squares estimation, is a cornerstone of applied mathematics. The gold-standard approach to solving the corresponding normal equations is applying the LQ decomposition (or its transpose, the QR decomposition). Here, we focus on {\em Bayesian linear regression} \cite [Sect.~3.5]{Bishop:06}, which is closely related to $\ell_2$ regularized least squares estimation.

Data $\{ (\vx{i}, y_i)\; |\; i=\rng{n}\}$, $y_i\in\R$, is fitted with $f(\vx{}) = \vx{}^T\vw{}$, a linear function in the weight vector $\vw{}\in\R^d$. The input vector $\vx{}\in\R^d$ consists of features, describing the data case. We assume Gaussian noise on the observed targets: $y_i\sim N(\vx{i}^T\vw{},\lambda_y)$. The feature map $\vx{}$ may depend on hyper-parameters $\vth{}$ (so that $\vx{} = \vx{}(\vth{})$). In Bayesian linear regression (BLR), we place a prior distribution on the weigths: $\vw{}\sim N(\vzero, \lambda_w\Id_d)$. If the feature vectors $\vx{i}$ are collected as columns in $\mxx{}\in\R^{d\times n}$, the model is given by prior and likelihood:
\[
  P(\vw{}) = N(\vzero, \lambda_w\Id_d),\quad P(\vy{}|\vw{}) = N(\vy{} |
  \mxx{}^T\vw{}, \lambda_y\Id_n),\quad \vy{}=[y_1,\dots,y_n]^T.
\]
We show how to learn hyper-parameters $\vth{}$, $\lambda_y$, $\lambda_w$ by maximizing the marginal likelihood
\[
  P(\vy{}) = N(\vy{} | \vzero,\mxsigma{y}),\quad \mxsigma{y} = \lambda_w
  \mxx{}^T\mxx{} + \lambda_y\Id_n.
\]
The learning criterion is
\[
  \phi = -\log P(\vy{}) \doteq \frac{1}2\left( \log|2\pi\mxsigma{y}| +
  \vy{}^T\mxsigma{y}^{-1}\vy{} \right).
\]
Define $\alpha = \lambda_w/\lambda_y$, $\mxm{} = \Id_d + \alpha\mxx{}\mxx{}^T$. If $\mxm{} = \mxl{}\mxl{}^T$ (Cholesky factorization), it can be shown that
\[
  \phi = \log|\mxl{}| + \frac{1}2\left( n\log(2\pi \lambda_y) + \lambda_y^{-1}
  \left( \|\vy{}\|^2 - \alpha\|\mxl{}^{-1}\mxx{}\vy{}\|^2 \right) \right).
\]
While we can proceed with {\tt linalg.potrf} as in \secref{ex-gp}, it is well known that least squares estimation is more robustly solved using the LQ factorization. In fact, if $\mxb{} = [\Id_d, \alpha^{1/2}\mxx{}] \in \R^{d\times (n+d)}$ and $\mxb{} = \mxl{}\mxq{}$, $\mxl{}\in\R^{d\times d}$ lower triangular and $\mxq{}\mxq{}^T = \Id_d$, then
\[
  \mxm{} = \mxb{}\mxb{}^T = \mxl{}\mxq{}\mxq{}^T\mxl{}^T = \mxl{}\mxl{}^T,
\]
so that $\mxl{}$ is equal to the Cholesky factor of $\mxm{}$.\footnote{
   This is true up to the signs of diagonal entries.}
Importantly, $\mxl{}$ is computed {\em without} first computing $\mxm{}$, which has numerical advantages, since $\mxb{}$ can have a smaller condition number than $\mxm{}$. We implemented the LQ factorization as differentiable operator {\tt linalg.gelqf}.  Here is Python code to create an MXNet symbolic expression for $\phi$:

\begin{minted}[frame=lines]{python}
import numpy as np
from mxnet.sym.linalg import gelqf, gemm2, syrk, potrf
from mxnet.sym import reshape, sqrt, concat, sum, square, log, abs

def bayeslinregr_criterion_symbol(feat_mat, targets, noise_var, prior_var,
        num_data, num_dim, use_lq=True, dtype=np.float32):
    alpha = prior_var / noise_var
    if use_lq:
        b1mat = mat_eye_symbol(num_dim, dtype)
        b2mat = feat_mat * sqrt(alpha)
        bmat = concat(*[b1mat, b2mat], dim=1)
        _, lmat = gelqf(bmat)
    else:
        mmat = syrk(feat_mat) * alpha + mat_eye_symbol(num_dim, dtype)
        lmat = potrf(mmat)
    yvec = reshape(targets, shape=(-1, 1))
    zvec = trsm(lmat, gemm2(feat_mat, yvec))
    sqnorm_y = sum(square(yvec))
    sqnorm_z = sum(square(zvec))
    logdet_l = sum(log(abs(extract_diag_symbol(lmat, num_data))))
    phi_part = (sqnorm_y - sqnorm_z * alpha) / noise_var
    return 0.5 * (phi_part + num_data * (np.log(2*np.pi) +
        log(noise_var))) + logdet_l
\end{minted}

Here, {\tt feat\_mat} represents the feature matrix $\mxx{}$, {\tt targets} the observed targets $\vy{}$, {\tt noise\_var} the noise variance $\lambda_y$, and {\tt prior\_var} the prior variance $\lambda_w$. The relevant operators are {\tt linalg.gelqf} (compute the LQ decomposition $\mxq{}$, $\mxl{}$ of matrix $\mxb{}$) and {\tt linalg.gemm2} (compute $\mxx{}\vy{}$), apart from {\tt linalg.syrk} and {\tt linalg.potrf} if $\mxl{}$ is determined by the Cholesky factorization of $\mxm{}$. See \secref{ex-gp} for other details, such as the computation of $\log|\mxl{}|$ ({\tt logdet\_l}).

Our example shows how hyper-parameter learning in a Bayesian linear regression model can be driven by MXNet. Once more, the real benefit lies in the ability to train hybrid models end-to-end, such as for example DNNs with BLR layers, or BLR with features maps given by DNNs. Our {\tt linalg} operators have been used to implement scalable Bayesian optimization and end-to-end learning for a hybrid model stacking Bayesian linear regression and a neural network \cite{Perrone:17}.

A Jupyter notebook that implements the Bayesian linear model based on the above symbolic expression is available at \url{http://github.com/ARCambridge/MXNet_linalg_examples}. In the notebook, we implement hyperparameter learning and prediction for the BLR model. The notebook is using the {\em GPy} framework (\url{https://github.com/SheffieldML/GPy}) for optimization and plotting, while all computations are done by MXNet executors. We also show how polynomial basis functions can be used as features. A demonstration is done on a toy 1D dataset.

\subsection{Kalman Filtering}
\label{sec:ex-kalman}

Gaussian linear dynamical systems (LDS) are among the most frequently used models for temporally varying data, with applications in tracking, robotics, control, or acoustic modelling. Inference is analytically tractable, via the Kalman filtering (and smoothing) algorithm. Different to time-varying neural network models, such as LSTMs or GRUs, the Kalman filter propagates distributions over latent states (instead of fixed values), thereby taking uncertainty and errors into account.

The Gaussian LDS is modelling a sequence of {\em observed} variables $[\vv{t}]$, where $t = 0, 1, \dots$ indexes time. It makes use of a sequence of {\em latent} states $\vh{t}$:
\[
  \vh{t}\sim N(\mxa{}\vh{t-1}, \mxsigma{h}),\quad \vv{t}\sim N(\mxb{}\vh{t},
  \mxsigma{v}).
\]
Moreover, $\vh{0}\sim N(\vmu{0},\mxsigma{0})$. The model parameters are transition matrices $\mxa{}$, $\mxb{}$, covariance matrices $\mxsigma{h}$, $\mxsigma{v}$, as well as $\vmu{0}, \mxsigma{0}$. They are learned by maximum likelihood estimation, typically driven by the expectation maximization algorithm, which in turn is powered by Kalman smoothing. In our use case, we fit the Gaussian LDS to a number of time series, and learn parameters by gradient-based optimization. The negative log likelihood is computed by Kalman filtering, which we implement in MXNet.

Full details are provided at \url{http://gluon.mxnet.io/chapter12_time-series/lds-scratch.html}.\footnote{
  The ``time series'' chapter of the interactive online book ``Deep Learning: The
  Straight Dope'', available at \url{http://gluon.mxnet.io/index.html}.}
Here, we sketch the part which requires our linear algebra operators. The filtering equations $(t-1)\to t$ are given by
\[
\begin{split}
  & \vmu{h} = \mxa{}\vf{t-1},\quad \vmu{v} = \mxb{}\vmu{h},\quad \mxsigma{h h}
  = \mxa{}\mxf{t-1}\mxa{}^T + \mxsigma{h}, \\
  & \mxsigma{v v} = \mxb{}\mxsigma{h h}\mxb{}^T + \mxsigma{v},\quad
  \mxk{t} = \mxsigma{h h}\mxb{}^T\mxsigma{v v}^{-1}, \\
  & \vf{t} = \vmu{h} + \mxk{t}(\vv{t} - \vmu{v}),\quad \mxf{t} =
  (\Id - \mxk{t}\mxb{}) \mxsigma{h h} (\Id - \mxk{t}\mxb{})^T +
  \mxk{t}\mxsigma{v}\mxk{t}.
\end{split}
\]
Here, $\mxk{t}$ is known as {\em Kalman gain matrix}. Here is the corresponding MXNet code snippet:

\begin{minted}[frame=lines]{python}
import numpy as np
import mxnet as mx
import mxnet.ndarray as nd
from mxnet.ndarray.linalg import gemm2, potrf, trsm

# ...

if t == 0:
    # At the first time step, use the prior
    mu_h = f_0
    S_hh = F_0
else:
    # Otherwise compute using update eqns.
    mu_h = gemm2(A, f_t)
    S_hh = gemm2(A, gemm2(F_t, A, transpose_b=1)) + S_h

# direct transcription of the update equations above
mu_v = gemm2(B, mu_h)
S_hh_x_B_t = gemm2(S_hh, B, transpose_b=1)
S_vv = gemm2(B, S_hh_x_B_t) + S_v

# use potrf to compute the Cholesky decomposition S_vv = L L^T
S_vv_chol = potrf(S_vv)

# K = (S_hh B^T) S_vv^{-1} = (S_hh B^T) L^{-T} L^{-1}
K = trsm(S_vv_chol, trsm(S_vv_chol, S_hh_x_B_t, rightside=1, transpose=1),
         rightside=1)

delta = v[t] - mu_v
f_t = mu_h + gemm2(K, delta)

ImKB = eye_h - gemm2(K, B)
F_t = gemm2(ImKB, gemm2(S_hh, ImKB, transpose_b=True))
              + gemm2(K, gemm2(S_v, K, transpose_b=True))
\end{minted}

Once more, we compute the Cholesky factor of $\mxsigma{v v}$ and use backsubstitution instead of computing its inverse, as the latter is less numerically stable.

Note how compared with the traditional parameter learning algorithm, things are much simpler here. Neither do we need expectation maximization, nor the backward pass of Kalman smoothing.\footnote{
  We gloss over some details here, such as the parameterization of the covariance
  matrices (which have to be kept positive definite).}
This is because the DLDS is taking care of the gradient computations. Once Gaussian LDS are integrated in MXNet, we can fit hybrid models to data. For example, the $\vv{t}$ could be modelled as sum of a deterministic LSTM and a Gaussian LDS, and parameters of the latter (such as $\mxa{}$ or $\mxsigma{v}$) could be parameterized by the LSTM as well. Even time series models with non-Gaussian likelihood, such as \cite{Seeger:16}, can be trained in MXNet.

\section{Linear Algebra Operators}
\label{sec:operators}

This section contains the main results of this work: forward and backward expressions for complex linear algebra operators such as Cholesky, LQ and symmetric eigen decomposition, along with details relevant for an efficient implementation. We present final expressions here, their derivations are collected in the Appendix. The following table summarizes the operators we contributed to MXNet.

\begin{center}
\begin{tabular}{|l|l|}
\hline
Operator & forward \\ \hline
$\mxl{} = \text{potrf}(\mxa{})$ & $\mxa{} = \mxl{}\mxl{}^T,\; l_{i j} = 0\; (i<j)$ \\
$(\mxq{}, \mxl{}) = \text{gelqf}(\mxa{})$ & $\mxa{} = \mxl{}\mxq{}$, $\mxq{}\mxq{}^T = \Id_m$ \\
$(\mxq{}, \vlam{}) = \text{syevd}(\mxa{})$ & $\mxa{} = \mxq{}^T (\text{diag}\vlam{}) \mxq{}$, $\mxq{}^T \mxq{} = \Id$ \\
\ifsvd
$(\mxu{}^T, \mxv{}, \vlam{}) = \text{gesvd}(\mxa{})$ & $\mxa{} = \mxu{}^T(\text{diag}\vlam{})\mxv{}$, $\mxu{}\mxu{}^T = \mxv{}\mxv{}^T = \Id_m$ \\
\fi
$\mxb{} = \text{trmm}(\mxl{}, \mxa{})$ & $\mxb{} = t(\mxl{})\mxa{}$, $\mxb{} = \mxa{} t(\mxl{})$ \\
$\mxb{} = \text{trsm}(\mxl{}, \mxa{})$ & $\mxb{} = t(\mxl{})^{-1}\mxa{}$, $\mxb{} = \mxa{} t(\mxl{})^{-1}$ \\
$\mxb{} = \text{potri}(\mxl{})$ & $\mxb{} = \mxa{}^{-1}$, $\mxa{}=\mxl{}\mxl{}^T$ \\
$\mxb{} = \text{syrk}(\mxa{})$ & $\mxb{} = t(\mxa{}) \bar{t}(\mxa{})$ \\
$\mxc{} = \text{gemm2}(\mxa{},\mxb{})$ & $\mxc{} = t_a(\mxa{}) t_b(\mxb{})$ \\
\hline
\end{tabular}
\end{center}

The first few are matrix factorizations, followed by support operators. Here, $\mxl{}$ is lower triangular, and $t(\mxx{}) = \mxx{}^T$ ($t=\text{true})$, $t(\mxx{}) = \mxx{}$ ($t=\text{false})$.

\subsection{Preliminaries}
\label{sec:oper-prelim}

{\em Reverse mode differentiation} (RMD) is a calculus for obtaining the gradient $\partial_{\vx{}} \phi(\vx{})$ of a scalar-valued function $\phi(\vx{})$. Here, $\vx{}$ is a vector, matrix, or multi-dimensional array. RMD can be seen as generalization of error backpropagation to differentiable directed {\em computation graphs}. Such bipartite graphs have variable nodes (multi-dimensional arrays) and operator nodes. The predecessor nodes of an operator are its inputs, the successor nodes are its outputs. A computation graph has a single scalar root node (no successors), which we call the {\em loss} and denote by $\phi$. The key strength of RMD is {\em modularity}: a simple graph traversal algorithm can compute the gradients of the loss w.r.t.\ any other variable node, if only {\em forward} and {\em backward} mapping are specified for every operator. In ML applications, certain input nodes (no predecessors) are parameters (weights, biases) to be learned, and this is done by gradient-based minimization of the loss, where RMD is used to compute the gradient.

Suppose we are given an operator $(\vc{},\vd{}) = \vf{}(\va{},\vb{})$, with two inputs and two outputs. The forward mapping is just $\vf{}$ itself, with inputs $\va{}$, $\vb{}$ and outputs $\vc{}$, $\vd{}$. For the backward mapping, we need some notation:
\[
   \bva{} = \partial_{\va{}}\phi,\quad \bvc{} = \partial_{\vc{}}\phi,\; \dots
\]
In its most general {\tt UseInOut} form, the backward mapping is
\[
  (\bva{}, \bvb{}) = \vf{\text{back}}(\bvc{},\bvd{},\va{},\vb{},\vc{},\vd{}).
\]
In other words, the backward mapping computes the {\em input gradients} $\bva{}$, $\bvb{}$, given the {\em output gradients} $\bvc{}$, $\bvd{}$, the inputs $\va{}$, $\vb{}$, and the outputs $\vc{}$, $\vd{}$. In some cases, the backward mapping requires a subset of these arguments only: $\vf{\text{back}}(\bvc{},\bvd{},\va{},\vb{})$ {\tt UseIn} or $\vf{\text{back}}(\bvc{},\bvd{},\vc{},\vd{})$ {\tt UseOut}.

RMD proceeds in two passes. In the forward pass, input nodes are bound to data, and all other nodes are computed by forward graph traversal. Note that even though we are formally only interested in the $\phi$ value, if gradients are required as well, we have to store values of all intermediate nodes. In the backward pass, the gradients $\bva{}$ are computed for all variable nodes $\va{}$ by backward graph traversal.

Highly efficient implementations of RMD, such as MXNet, aim to save memory whenever possible. After all, the main bottleneck of graphics processing units (GPUs) is limited on-board memory. If no gradients are required, there is no need to store intermediate node values, and downstream nodes can be overwritten by upstream ones. Even in forward-backward mode, memory usage can be optimized by rearranging the graph traversal. For our work on operators, it is important that we can signal MXNet that certain outputs can overwrite certain inputs. For example, the code of $\vf{}$ may allow $\vc{}$ to overwrite $\va{}$, or in $\vf{\text{back}}$, $\bva{}$ may overwrite $\bvc{}$. Another important concept is {\em in-place computation}. Whenever possible, our forward and backward implementations will employ steps where outputs overwrite inputs not needed anymore, with the goal of using as little temporary memory as possible (most of our operators do not require temporary memory at all).

Finally, note that all our operators support batch operations. Instead of 2D matrices, they also accept 3D tensors. The first dimension is then iterated over. We refer to this variant as {\em batch mode} below.

\subsubsection*{Simple Matrix Functions}

The Hadamard (pointwise) product of two matrices of the same shape is defined as
\[
  [\mxa{}\circ\mxb{}]_{i j} = a_{i j} b_{i j}.
\]
$\sym(\mxx{})$ symmetrizes a square matrix:
\[
  \sym(\mxx{}) = \frac{1}2\left( \mxx{} + \mxx{}^T \right).
\]
$\tril(\mxx{})$ extracts the lower triangle (and diagonal) from a square matrix:
\[
  [\tril(\mxx{})]_{i j} = x_{i j} \Ind{i \ge j}.
\]
$\text{copyltu}(\mxx{})$ for a square matrix $\mxx{}$ generates a symmetric matrix by copying the lower triangle to the upper triangle:
\[
  [\text{copyltu}(\mxx{})]_{i j} = x_{\max(i,j), \min(i,j)}.
\]

\subsubsection*{BLAS and LAPACK}

Our operators are implemented by calling tuned numerical linear algebra code, adhering to the BLAS and LAPACK APIs. For our MXNet implementation, we adopt their naming scheme. For example, the Cholesky decomposition is {\tt dpotrf} in {\tt float64}, {\tt spotrf} in {\tt float32} in LAPACK, while our operator is called {\tt linalg.potrf} (we drop the prefix d or s, since both types are supported via the same operator). Implementations details are given in \secref{mxnet}. In short, we make use of the existing BLAS dependency for MXNet on CPUs, while using CUBLAS and CUSolver for GPU support.\footnote{
  We are adding LAPACK wrappers on demand. For CPUs, a BLAS dependency
  typically includes LAPACK.}

We only use the simplest flat storage format for matrices in BLAS/LAPACK, avoiding compressed formats for triangular matrices. A lower triangular matrix $\mxl{}$ has $l_{i j} = 0$ for all $i<j$ (note that the diagonal is non-zero in general). A symmetric matrix $\mxa{}$ has $a_{i j} = a_{j i}$ for all $i, j$. While BLAS/LAPACK represent symmetric matrices by triangular ones, and only access respective triangles, our MXNet implementation always uses a full square matrix. We note that while it is possible to support compressed matrix formats, and we may do so in the future, such a practice runs against the spirit of DLDS. Once outputs of an operator are not just unconstrained multi-dimensional arrays, it cannot be linked with any other operator.

When using BLAS/LAPACK with MXNet, one additional complication has to be dealt with. BLAS/LAPACK represents matrices in {\em column-major} storage:
\[
  \left[ \begin{array}{cc}
    a_{1 1} & a_{1 2} \\
    a_{2 1} & a_{2 2}
  \end{array}\right]\quad \Rightarrow \quad
  \left[ a_{1 1}\, a_{2 1}\, a_{1 2}\, a_{2 2} \right].
\]
MXNet follows NumPy in using {\em row-major} storage:
\[
  \left[ \begin{array}{cc}
    a_{1 1} & a_{1 2} \\
    a_{2 1} & a_{2 2}
  \end{array}\right]\quad \Rightarrow \quad
  \left[ a_{1 1}\, a_{1 2}\, a_{2 1}\, a_{2 2} \right].
\]
While certain wrappers of BLAS/LAPACK seem to support row-major storage, this often comes at the cost of hidden transpositions and extra memory. Our solution avoids any such conversions. We implement an operator in MXNet by calling respective operators on tranposes internally. For example, we offer the LQ decomposition, but internally call LAPACK code for the QR decomposition (\secref{oper-lq}).

\subsection{Cholesky Decomposition. Related Operators}
\label{sec:oper-chol}

Derivations of expressions are given in the Appendix.
Given a symmetric, positive definite matrix $\mxa{}$, its Cholesky factor $\mxl{}$ is lower triangular with positive diagonal, such that $\mxa{} = \mxl{}\mxl{}^T$.
{\em forward}:
\[
  \mxl{} = \text{potrf}(\mxa{}),\quad \mxa{} = \mxl{}\mxl{}^T.
\]
Memory: $\mxl{}$ can overwrite $\mxa{}$. No extra memory.

{\em backward}:
Inputs: $\bmxl{}$, $\mxl{}$ ({\tt UseOut}).
\[
  \bmxa{} = \frac{1}2 \mxl{}^{-T} \text{copyltu}(\mxl{}^T\bmxl{}) \mxl{}^{-1}
\]
Here, $\text{copyltu}$ is defined in \secref{oper-prelim}. This expression has been given in \cite{Murray:16}. \\
Memory: $\bmxa{}$ can overwrite $\bmxl{}$. No extra memory.

It is important to note that the computation of $\bmxa{}$ is in-place (no extra memory is needed). We start with copying $\bmxl{}$ to $\bmxa{}$. Both the multiplication with $\mxl{}^T$ and the backsubstitution operators are in-place BLAS functions.

In order to make {\tt potrf} useful, we implemented a number of additional operators in MXNet. These are listed in the sequel.

\subsubsection*{gemm2}

Matrix-matrix multiplication. {\em forward}:
\[
  \mxc{} = \text{gemm2}(\mxa{},\mxb{}; t_a, t_b, \alpha) = \alpha t_a(\mxa{})
  \cdot t_b(\mxb{})
\]
Here, $t(\mxx{}) = \mxx{}$ if $t = \text{false}$, $t(\mxx{}) = \mxx{}^T$ if $t = \text{true}$. \\
No extra memory required.

{\em backward}:
Inputs: $\bmxc{}$, $\mxa{}$, $\mxb{}$ ({\tt UseIn}).
\[
  \bmxa{} = \left\{ \begin{array}{ll}
    \text{gemm2}(\bmxc{},\mxb{}; t_a, \bar{t_b}, \alpha),\quad & t_a=\text{false} \\
    \text{gemm2}(\mxb{},\bmxc{}; t_b, t_a, \alpha),\quad & t_a=\text{true}
  \end{array}\right\}, \quad \bmxb{} = \left\{ \begin{array}{ll}
    \text{gemm2}(\mxa{}, \bmxc{}; \bar{t_a}, t_b, \alpha),\quad & t_b=\text{false} \\
    \text{gemm2}(\bmxc{}, \mxa{}; t_b, t_a, \alpha),\quad & t_b=\text{true}
  \end{array}\right\}
\]
Memory: $\bmxb{}$ can overwrite $\mxb{}$, but only if $\bmxa{}$ is computed before $\bmxb{}$. No extra memory required.

Note that the general matrix-matrix multiplication does not allow for in-place computation. It is maybe surprising that such an operator has to be added to MXNet, given that most neural networks contain densely connected layers. The reason is that other MXNet operators do not consistently support {\tt float64}, and may also be unwieldy to use.

\subsubsection*{trmm}

Multiplication with lower triangular matrix (in-place). There are four different cases. {\em forward}:
\[
\begin{split}
  \mxb{} = \text{trmm}(\mxl{},\mxa{}; t_l, \text{rightside} = \text{false}) & = t_l(\mxl{}) \mxa{}, \\
  \mxb{} = \text{trmm}(\mxl{},\mxa{}; t_l, \text{rightside} = \text{true}) & = \mxa{} t_l(\mxl{}) \\
\end{split}
\]
$\mxl{}$ is lower triangular. $t(\mxx{}) = \mxx{}$ if $t = \text{false}$, $t(\mxx{}) = \mxx{}^T$ if $t = \text{true}$. \\
Memory: $\mxb{}$ can overwrite $\mxa{}$. No extra memory.

{\em backward}:
Inputs: $\bmxb{}$, $\mxl{}$, $\mxa{}$ ({\tt UseIn}).
\[
\begin{split}
  \mxb{}=\mxl{}\mxa{}\quad & \Rightarrow\quad \bmxa{}=\mxl{}^{T}\bmxb{},
  \; \bmxl{} = \tril\left( \bmxb{}\mxa{}^T \right), \\
  \mxb{}=\mxl{}^T\mxa{}\quad & \Rightarrow\quad \bmxa{}=\mxl{}\bmxb{},
  \; \bmxl{} = \tril\left( \mxa{}\bmxb{}{}^T \right), \\
  \mxb{}=\mxa{}\mxl{}\quad & \Rightarrow\quad \bmxa{}=\bmxb{}\mxl{}^{T},
  \; \bmxl{} = \tril\left( \mxa{}^T\bmxb{} \right), \\
  \mxb{}=\mxa{}\mxl{}^T\quad & \Rightarrow\quad \bmxa{}=\bmxb{}\mxl{},
  \; \bmxl{} = \tril\left( \bmxb{}{}^T\mxa{} \right) \\
\end{split}
\]
Memory: $\bmxa{}$ can overwrite $\bmxb{}$, given that $\bmxl{}$ is computed first. No extra memory.

Note that not only is {\em trsm} faster than {\em gemm2}, it is also in-place, so that $\mxb{}$ can overwrite $\mxa{}$.

\subsubsection*{trsm}

Backsubstitution with lower triangular matrix (in-place). There are four different cases. {\em forward}:
\[
\begin{split}
  \mxb{} = \text{trsm}(\mxl{},\mxa{}; t_l, \text{rightside} = \text{false}) & = t_l(\mxl{})^{-1} \mxa{}, \\
  \mxb{} = \text{trsm}(\mxl{},\mxa{}; t_l, \text{rightside} = \text{true}) & = \mxa{} t_l(\mxl{})^{-1} \\
\end{split}
\]
$\mxl{}$ is lower triangular. $t(\mxx{}) = \mxx{}$ if $t = \text{false}$, $t(\mxx{}) = \mxx{}^T$ if $t = \text{true}$. \\
Memory: $\mxb{}$ can overwrite $\mxa{}$. No extra memory.

{\em backward}:
Inputs: $\bmxb{}$, $\mxl{}$, $\mxa{}$, $\mxb{}$ ({\tt UseInOut}).
\[
\begin{split}
  \mxb{}=\mxl{}^{-1}\mxa{}\quad & \Rightarrow\quad \bmxa{}=\mxl{}^{-T}\bmxb{},
  \; \bmxl{} = -\tril\left( \bmxa{}\mxb{}^T \right), \\
  \mxb{}=\mxl{}^{-T}\mxa{}\quad & \Rightarrow\quad \bmxa{}=\mxl{}^{-1}\bmxb{},
  \; \bmxl{} = -\tril\left( \mxb{}\bmxa{}{}^T \right), \\
  \mxb{}=\mxa{}\mxl{}^{-1}\quad & \Rightarrow\quad \bmxa{}=\bmxb{}\mxl{}^{-T},
  \; \bmxl{} = -\tril\left( \mxb{}^T\bmxa{} \right), \\
  \mxb{}=\mxa{}\mxl{}^{-T}\quad & \Rightarrow\quad \bmxa{}=\bmxb{}\mxl{}^{-1},
  \; \bmxl{} = -\tril\left( \bmxa{}{}^T\mxb{} \right)
\end{split}
\]
Memory: $\bmxa{}$ can overwrite $\bmxb{}$ or $\mxa{}$. $\bmxl{}$ can overwrite $\mxl{}$. Here, $\bmxa{}$ has to be computed first. No extra memory.

\subsubsection*{syrk}

Multiplication of matrix with own transpose. {\em forward}:
\[
  \mxb{} = \text{syrk}(\mxa{}; t_a, \alpha) = \alpha t_a(\mxa{})\cdot
  \bar{t_a}(\mxa{}).
\]
Here, $t(\mxx{}) = \mxx{}$ if $t = \text{false}$, $t(\mxx{}) = \mxx{}^T$ if $t = \text{true}$. \\
No extra memory required.

{\em backward}:
Inputs: $\bmxb{}$, $\mxa{}$ ({\tt UseIn}). Note that $\bmxb{}$ need not be symmetric.
\[
\begin{split}
  \mxb{} = \alpha\mxa{}\mxa{}^T\quad & \Rightarrow\quad
  \bmxa{} = 2\alpha(\sym\bmxb{})\mxa{}, \\
  \mxb{} = \alpha\mxa{}^T\mxa{}\quad & \Rightarrow\quad
  \bmxa{} = 2\alpha\mxa{}(\sym\bmxb{})
\end{split}
\]
No extra memory required.

\subsubsection*{potri}

Computation of inverse $\mxa{}^{-1}$, given the Cholesky factor $\mxl{}$ of $\mxa{}$, where $\mxa{}$ is symmetric, positive definite. $\mxl{}$ lower triangular with positive diagonal. {\em forward}:
\[
  \mxb{} = \text{potri}(\mxl{}) = \mxa{}^{-1},\quad \mxa{} = \mxl{}\mxl{}^T.
\]
Memory: $\mxb{}$ can overwrite $\mxl{}$. No extra memory required.

{\em backward}:
Inputs: $\bmxb{}$, $\mxl{}$, $\mxb{}$ ({\tt UseInOut}). Note that $\bmxb{}$ need not be symmetric:
\[
  \bmxl{} = -2 \tril\left( \mxb{}\sym(\bmxb{})\mxl{}^{-T} \right).
\]
No extra memory required.

This operator is needed only in special circumstances. For example, some algorithms require the diagonal of the inverse of $\mxa{}$. However, most expressions in ML algorithms can be computed with the Cholesky factor alone, together with {\tt trmm} and {\tt trsm}. For example, the $\phi$ expression in \secref{ex-gp} contains $\mxa{}^{-1}$, but the code does not use {\tt potri}. As a general rule, explicit computation of the inverse $\mxa{}^{-1}$ should be avoided whenever possible: inversion is subject to a lot more numerical error than Cholesky factorization alone. It is also at least twice as expensive to compute.

\subsection{LQ Decomposition}
\label{sec:oper-lq}

Derivations of expressions are given in the Appendix.
Suppose that $\mxa{}\in\R^{m\times n}$, where $m\le n$. The LQ decomposition is
\[
  \mxa{} = \mxl{}\mxq{},\quad \mxq{}\mxq{}^T = \Id_m,\quad l_{i j} = 0\; (i<j).
\]
Here, $\mxq{}\in\R^{m\times n}$ is row-orthonormal, and $\mxl{}\in\R^{m\times m}$ is lower triangular with non-zero diagonal. We require that $\mxa{}$ has full rank: $\rank\mxa{} = m$.\footnote{
  The operator does not have derivatives at rank-deficient $\mxa{}$.}
{\em forward}:
\[
  (\mxq{}, \mxl{}) = \text{gelqf}(\mxa{}),\quad \mxa{} = \mxl{}\mxq{},\quad
  \mxq{}\mxq{}^T = \Id_m.
\]
Memory: $\mxq{}$ can overwrite $\mxa{}$. Additional workspace memory is required.

{\em backward}:
Inputs: $\bmxq{}$, $\bmxl{}$, $\mxq{}$, $\mxl{}$ ({\tt UseOut}). Note that $\bmxl{}$ need not be lower triangular:
\[
  \bmxa{} = \mxl{}^{-T} \left( \bmxq{} + \text{copyltu}(\mxm{})\mxq{} \right),
  \quad \mxm{} = \mxl{}^T\bmxl{} - \bmxq{}\mxq{}^T.
\]
Here, $\text{copyltu}$ is defined in \secref{oper-prelim}. This expression is novel to the best of our knowledge. \\
Memory: $\bmxa{}$ can overwrite $\bmxq{}$. One temporary $\R^{m\times m}$ matrix is required.

\subsubsection*{Implementation Details}

LAPACK code for the LQ factorization is more complicated than for Cholesky. First, we have to call two LAPACK routines: {\tt gelqf}, followed by {\tt orglq}. The first returns $\mxq{}$ in an internal storage format, which is converted to matrix form by the second. Second, both routines require working space, the amount of which depends on $n$, $m$ and internal parameters. The amount of working space has to be requested by a workspace query. Our implementation calls these queries and allocates workspace space accordingly. In batch mode, workspace is allocated once only, then used for all items of a batch.

The column versus row-major issue was noted in \secref{oper-prelim}. This means that internally, we compute the QR decomposition of $\mxa{}^T$, calling LAPACK {\tt geqrf} and {\tt orgqr}. For ML applications, we can always use the LQ factorization in place of the QR factorization. Given that LAPACK offers both LQ and QR factorization, why do we not offer both as well? It turns out that the current version of CUSolver only contains the QR factorization (in column-major).

\subsection{Symmetric Eigendecomposition}
\label{sec:oper-eigen}

Derivations of expressions are given in the Appendix.
Given a symmetric matrix $\mxa{}\in\R^{n\times n}$, its eigendecomposition is
\[
  \mxa{} = \mxu{}^T(\diag\vlam{})\mxu{},\quad \mxu{}^T\mxu{} = \Id.
\]
Here, $\mxu{}$ is orthonormal, and the eigenvalues are ascending: $\lambda_1\le\lambda_2\le\dots\le\lambda_n$. The rows of $\mxu{}$ are the eigenvectors corresponding to eigenvalues. In order for derivatives to be defined, all eigenvalues of $\mxa{}$ must be distinct. Our implementation still works if $\vlam{}$ has very small or zero eigengaps, but in this case, {\em backward} is approximate. Note that if the entries of $\mxa{}$ are drawn i.i.d.\ at random (subject to being symmetric), the eigenvalues are distinct with probability one. However, the user has to be careful not to apply {\tt syevd} to matrices such as $\Id + \mxx{}\mxx{}^T$, which have multiple eigenvalues by design. {\em forward}:
\[
  (\mxu{}, \vlam{}) = \text{syevd}(\mxa{}),\quad \mxa{} =
  \mxu{}^T(\diag\vlam{})\mxu{},\quad \mxu{}^T\mxu{} = \Id.
\]
Memory: $\mxu{}$ can overwrite $\mxa{}$. Additional workspace memory is required.

{\em backward}:
Inputs: $\bmxu{}$, $\bvlam{}$, $\mxu{}$, $\vlam{}$ ({\tt UseOut}):
\[
  \bmxa{} = \mxu{}^T \left( \sym(\bmxu{}\mxu{}^T \circ\mxf{}) + \bmxlam{}
  \right) \mxu{},\quad F_{i j} = \frac{\Ind{i\ne j}}{h(\lambda_i - \lambda_j)},\quad
  h(t) = \max(|t|,\eps) \sgn(t),
\]
and $\eps>0$ is a small scalar. Note that using $h(\cdot)$ amounts to an approximation. The correct expression has $h(t) = t$, resulting in numerical errors for vanishing eigengaps. This expression is given in \cite{Giles:08}. \\
Memory: $\bmxa{}$ can overwrite $\bmxu{}$. One temporary matrix of the shape of $\mxa{}$ is needed.

If $\mxy{} = \sym(\mxx{}\circ\mxf{})$, then $y_{i i} = 0$, and for $i>j$:
\[
  y_{i j} = \frac{x_{i j} - x_{j i}}{2 h(\lambda_i - \lambda_j)} =
  \frac{x_{i j} - x_{j i}}{2 \max(\lambda_i - \lambda_j, \eps)},\quad \text{since}\;
  \lambda_i\ge\lambda_j.
\]

One additional issue has to be dealt with: the signs of eigenvectors are not well defined. In other words, if $\mxu{}$ collects eigenvectors of $\mxa{}$, so does $(\diag\vsigma{})\mxu{}$, where $\sigma_i\in\{\pm 1\}$. The backward expression is invariant to sign changes only if $\bmxu{}$ is replaced by $(\diag\vsigma{})\bmxu{}$ as well. Denote the $i$-th row of $\mxu{}$ by $\vu{i}$. We need a deterministic rule to decide upon the sign of each $\vu{i}$. Any rule which splits the unit sphere in half will do. Our implementation determines $k_i = \argmax_k |u_{i k}|$. If $u_{i k_i} < 0$, then $\vu{i}$ is replaced by $-\vu{i}$. In case of a tie, the sign is determined by the smaller index $k_i$. Note that any such rule does give rise to further (isolated) inputs $\mxa{}$ at which derivatives are not well defined.

\subsubsection*{Implementation Details}

The standard LAPACK routine for symmetric eigendecomposition is {\tt syevd}. Note that there is a variant {\tt syevr}, which is reported to be faster and to require less memory. On the flip side, calling {\tt syevr} is slightly more involved, and $\mxu{}$ cannot overwrite $\mxa{}$. Also, this routine is not currently implemented in CUSolver. Our current implementation uses {\tt syevd} on both CPU and GPU, leaving the support of {\tt syevr} on the CPU for future work.

The column versus row-major issue was noted in \secref{oper-prelim}. This means that internally, we compute the factorization $\mxa{} = \mxu{}(\diag\vlam{})\mxu{}^T$, where eigenvectors are columns of $\mxu{}$.

\ifsvd
\subsection{Singular Value Decomposition}
\label{sec:oper-svd}

Derivations of expressions are given in the Appendix.
Let $\mxa{}\in\R^{m\times n}$, where $m\le n$. The singular value decomposition (SVD) is given by
\[
  \mxa{} = \mxu{}^T\mxlam{}\mxv{},\quad \mxlam{} = \diag\vlam{}\in
  \R^{m\times m},\; \mxu{}\in\R^{m\times m},\; \mxv{}\in\R^{m\times n}.
\]
Here, $\mxu{}$ and $\mxv{}$ are orthonormal:
\[
  \mxu{}\mxu{}^T = \mxu{}^T\mxu{} = \Id_m,\quad \mxv{}\mxv{}^T = \Id_m.
\]
The singular values $\lambda_i$ are non-zero and ascending:
\[
  0\le \lambda_1\le\lambda_2\le\dots\le\lambda_m.
\]
In order for derivatives to be well-defined in this ``thin SVD'' case, the singular values must be positive and distinct:
\[
  0 < \lambda_1 < \lambda_2 < \dots < \lambda_m.
\]
In particular, the matrix $\mxa{}$ must have full rank $m$. Our implementation does not fail if some singular values are equal, but the backward expression is wrong then (the exact expression being undefined). {\em forward}:
\[
  (\mxu{}, \vlam{}, \mxv{}) = \text{gesvd}(\mxa{}),\quad \mxa{} =
  \mxu{}^T(\diag\vlam{})\mxv{},\quad \mxu{}\mxu{}^T = \mxv{}\mxv{}^T = \Id_m.
\]
Memory: $\mxv{}$ can overwrite $\mxa{}$. Additional workspace memory is required.

{\em backward}:
Inputs: $\bmxu{}$, $\bmxv{}$, $\bvlam{}$, $\mxu{}$, $\mxv{}$, $\vlam{}$, ({\tt UseOut}):
\[
\begin{split}
  \bmxa{} & = \mxu{}^T\left( \mxg{2}\mxv{} + \mxlam{}^{-1}\bmxv{} \right),\quad
  \mxg{2} = \bmxlam{} + 2\sym(\mxg{1}\circ\mxe{})\mxlam{} - (\mxlam{}^{-1}
  \bmxv{}\mxv{}^T \circ\Id), \\
  \mxg{1} & = \bmxu{}\mxu{}^T + \mxlam{}^{-1}\bmxv{}\mxv{}^T\mxlam{},\quad
  E_{i,j} = \frac{\Ind{i\ne j}}{h(\lambda_i - \lambda_j) h(\lambda_i + \lambda_j)},
  \quad h(t) = \max(|t|,\eps) \sgn(t),
\end{split}
\]
and $\eps>0$ is a small scalar. Note that using $h(\cdot)$ amounts to an approximation. The correct expression has $h(t) = t$, resulting in numerical errors for vanishing gaps between singular values. A related, but different expression is given in \cite{Giles:08}. This expression holds for the full SVD case, where $\mxv{}\in\R^{n\times n}$, and $\mxlam{}\in\R^{m\times n}$ is padded with zero columns. The full SVD case is not useful in the machine learning context: $\mxv{}\in\R^{n\times n}$ is potentially very large, its last $n-m$ columns are not used. Finally, a derivation by James Townsend at \url{https://j-towns.github.io/papers/svd-derivative.pdf} seems equivalent to ours here, but implementation details are not provided. \\
Memory: $\bmxa{}$ can overwrite $\bmxv{}$. One temporary $\R^{m\times m}$ matrix is needed. As detailed in the Appendix, the final left-multiplication with $\mxu{}^T$ can be done in-place  by chunking the $m\times n$ matrix into $m\times m$ blocks.

If $\mxy{} = 2\sym(\mxx{}\circ\mxe{})$, then $y_{i i} = 0$, and for $i>j$:
\[
  y_{i j} = \frac{x_{i j} - x_{j i}}{h(\lambda_i - \lambda_j) h(\lambda_i +
  \lambda_j)},\quad h(t) = \max(t, \eps),\quad \text{since}\; \lambda_i >
  \lambda_j.
\]
Moreover, we use the same mechanism as for {\tt syevd} in order to choose the sign of rows of $\mxu{}$, $\mxv{}$. This is mainly useful in order to allow for finite difference testing.

\subsubsection*{Implementation Details}

We use the LAPACK routine {\tt gesvd}. The column versus row-major issue was noted in \secref{oper-prelim}. This means that internally, we compute the SVD $\mxa{}^T = \mxv{}^T\mxlam{}\mxu{}$. Here, {\tt u} maps to $\mxv{}^T$, {\tt vt} to $\mxu{}^T$. Therefore, we set {\tt jobu = 'O'} (so that $\mxv{}^T$ overwrites $\mxa{}^T$) and {\tt jobvt = 'A'}.

\fi

\section{Implementation in MXNet}
\label{sec:mxnet}

In this section, we provide details on our implementation of differentiable linear algebra operators in the MXNet DLDS.\footnote{
  \url{https://github.com/apache/incubator-mxnet}}
Different to TensorFlow, which depends on Eigen (see \secref{rel-software}), MXNet did not support linear algebra expressions beyond matrix-matrix multiplication before our extension. We decided to implement the new operators directly in the core C++ kernel of MXNet, instead of going for a Python based extension, as is the case for TensorFlow.\footnote{
  This may be due to the fact that open source extensions to TensorFlow are
  limited to its Python API, while MXNet invites open source contributions to its
  core just as well.}
Doing so enabled us to
\begin{itemize}
 \item ensure maximum performance and efficiency of the operator implementations
 \item minimize the memory footprint of the operators, by exploiting all opportunities for in-place computations and memory re-usage (as the C++ core of MXNet allows fine-grained control over memory management)
\item leverage the MXNet intrinsic mechanisms that allow contributors to write a single C++ implementation, from which language bindings for all supported languages (Python, Scala, Julia, Perl, R) are automatically generated, both for imperative and declarative programming models
\end{itemize}

The implementation is broken up into several MXNet-internal layers of abstraction. The goal was to provide an infrastructure for linear algebra operations within MXNet that is efficient and extendable. We anticipate that there will be demand for adding further functionality in the future, and want to lay out a specific software pattern for future contributors. This makes it easier to add new operators for linear algebra, and at the same time helps maintaining a consistent code structure.

For CPU support, we integrated the LAPACK library into MXNet in addition to the already imported BLAS library. Integration is done by interfacing directly to the Fortran code. We decided against using existing C wrappers of LAPACK, such as {\em lapacke}. These are not consistently available across all MXNet-supported platforms, and they tend to be implemented inefficiently, relying on unnecessary transpositions and temporary copies in order to bridge between row-major and column-major storage. In contrast, our wrappers are consistently in-place. They may modify arguments (such as flipping flags for upper or lower triangle, or transposition), the input operand order, and sometimes even the type of the requested computation (such as calling the LAPACK QR factorization on the transpose matrix, see \secref{oper-lq}). As a result, the desired computation in row-major layout (MXNet) is translated into an equivalent computation based on column-major layout (LAPACK, BLAS) without any performance or memory penalty.

We decided against wrapping the entire BLAS and LAPACK APIs in one shot, but only imported functions which were required in the scope of this project. We feel it is best to integrate functionality on demand, each time carefully reducing computation overhead, rather than mechanically integrating the full library.

For GPU support, we integrated the cuSOLVER library (\url{https://developer.nvidia.com/cusolver}) into MXNet, in addition to the already imported cuBLAS library (\url{https://developer.nvidia.com/cublas}). cuSOLVER implements a subset of the LAPACK API and is specifically tuned for NVidia GPUs. The subset covers all functions needed for our operators, except {\em potri} (inverse from Cholesky factor). We implemented {\em potri} for GPUs by two calls to {\em trsm}. Since cuSOLVER assumes column-major storage, we applied the techniques described above to map to the row-major layout required for MXNet.

The second abstraction layer implements a unified tensor-based C++ interface for BLAS/LAPACK functions. It provides overloaded functions that execute the same linear algebra operations on CPU or GPU and supports processing individual matrices as well as batch mode operations that involve multiple matrices having the same dimensions. This layer hides all aspects of matrix memory layout, differences in function signatures between CPU and GPU versions, and details of parallel processing from the caller. It also hides the fact that {\tt float32} and {\tt float64} BLAS/LAPACK functions have different names. The currently supported operators are detailed in \secref{operators}.

MXNet internally follows a program pattern, where input and output tensors of an operator are located either on the CPU or one of multiple GPUs. The execution context of an operation can therefore be deduced from its arguments, it does not have to be specified. This convenient pattern allows to write most code in a generic fashion, independent of the execution device.

The third abstraction layer is built entirely on top of the second one, and is therefore agnostic to CPU or GPU usage, as well as data type. It provides C++ methods for doing efficient forward and backward computations of the linear algebra operators. The forward computation is usually a direct call to the second layer methods, while the backward computation comprises of a series of basic linear algebra operations. As noted in \secref{operators}, most backward computations are entirely in-place. This is possible because the BLAS/LAPACK API supports in-place whenever possible, and so does our second layer.

Based on the third abstraction layer, high level operators have been added to MXNet that are accessible to the user in a multitude of different languages. Creating such operators involves packaging of the linear algebra methods into data structures that are understood by core MXNet components, such as the scheduler and the memory manager. MXNet provides a convenient registry mechanism for accomplishing this. It only requires to register C++ functions for the forward and backward pass that should be executed whenever the operator is invoked as part of a user program. Alongside with the entry points to these functions, additional attributes can be supplied that control memory allocation and memory re-usage, parameter passing, type checking and even documentation. Leveraging the registry and using generic templated wrappers around the various linear algebra methods, all linear algebra operators visible to the user are realized in a concise and unified way.

\section{Experiments}
\label{sec:exper}

In this section, we present results of experiments in order to understand the benefits of implementing advanced machine learning methods in MXNet, using our {\tt linalg} extension, compared to using state of the art codes. We are mainly interested in faster execution, both on CPU and GPU, and in simplification of implementations. While the former can be quantified in runtime comparisons, the latter can be understood from code snippets in \secref{examples} and notebooks provided in the supplemental material.

\subsection{Operator Runtime Comparison with TensorFlow}
\label{sec:exper-tf}

As noted in \secref{relwork}, TensorFlow provides differentiable operators for the same matrix factorizations as our {\tt linalg.potrf}, {\tt linalg.gelqf}, and {\tt linalg.syevd}. We compared running times of forward and backward operators for all three cases.

We observed that on Intel CPUs, the Intel-MKL implementation of BLAS/LAPACK provides the best performance.\footnote{
  It is important to note that this BLAS/LAPACK library is different from what is
  known as MKL-ML, they overlap on dense matrix-matrix multiplication only.}
Therefore, we linked MXNet against Intel-MKL for CPU, and against NVidia's cuBLAS/cuSOLVER libraries for GPU processing. For the TensorFlow CPU experiments, we built against the master branch as of Jan 20 2018 with MKL support. The GPU experiments were done using {\tt tensorflow-gpu==1.4.1}, since building from source with proper GPU support did not work for us. While MXNet experiments ran efficiently out of the box, we had to apply a variety of optimization tricks for data loading and unloading, so that TensorFlow code gave optimal performance. CPU tests were run on an AWS EC2 {\tt c4.8xlarge} host (36 virtual cores), GPU tests on a {\tt p3.2xlarge} instance using a Tesla v100 GPU, CUDA 9, and CUDNN 4.

\begin{figure*}
\minipage{0.49\textwidth}
  \includegraphics[width=\linewidth]{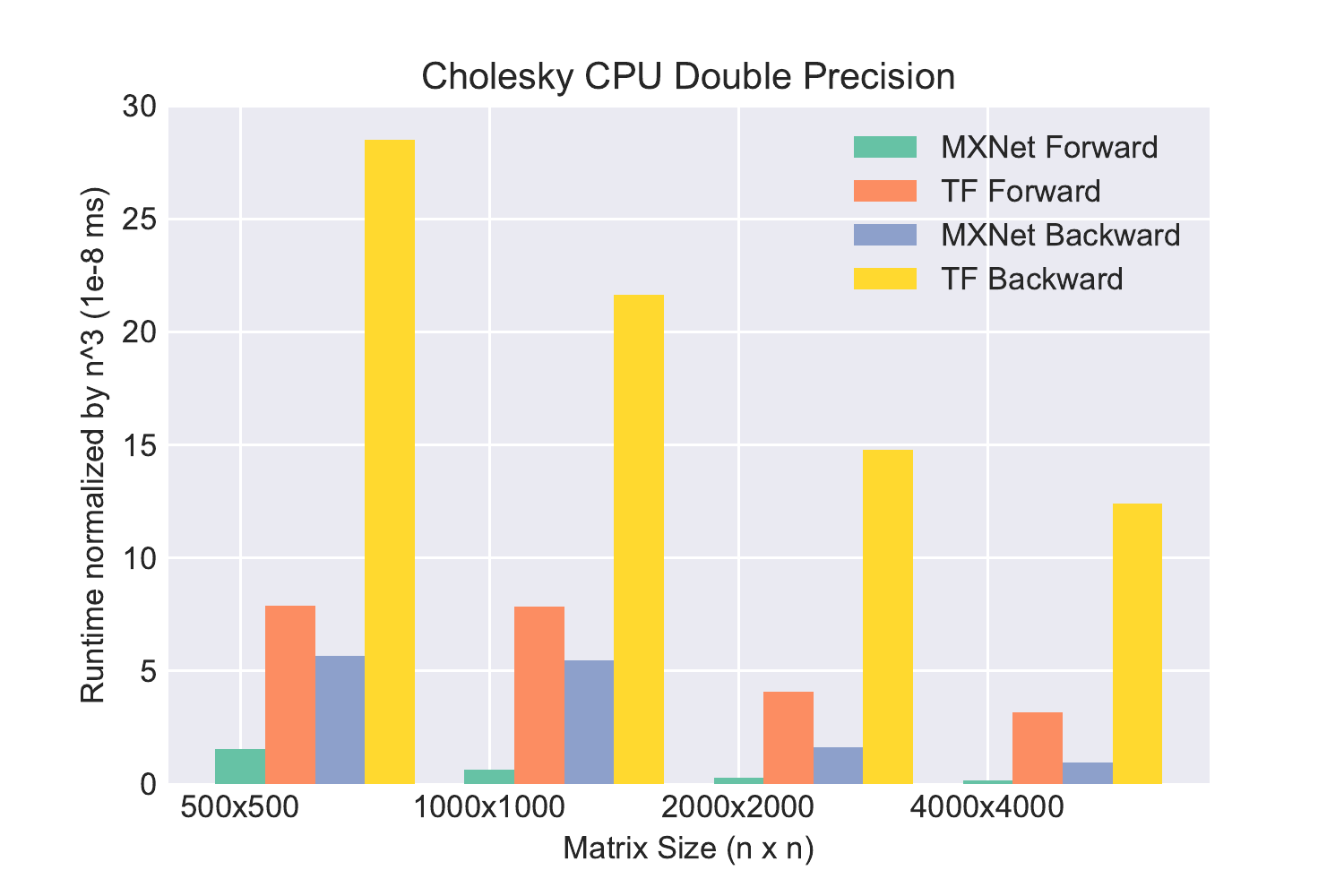}
\endminipage\hfill
\minipage{0.49\textwidth}
  \includegraphics[width=\linewidth]{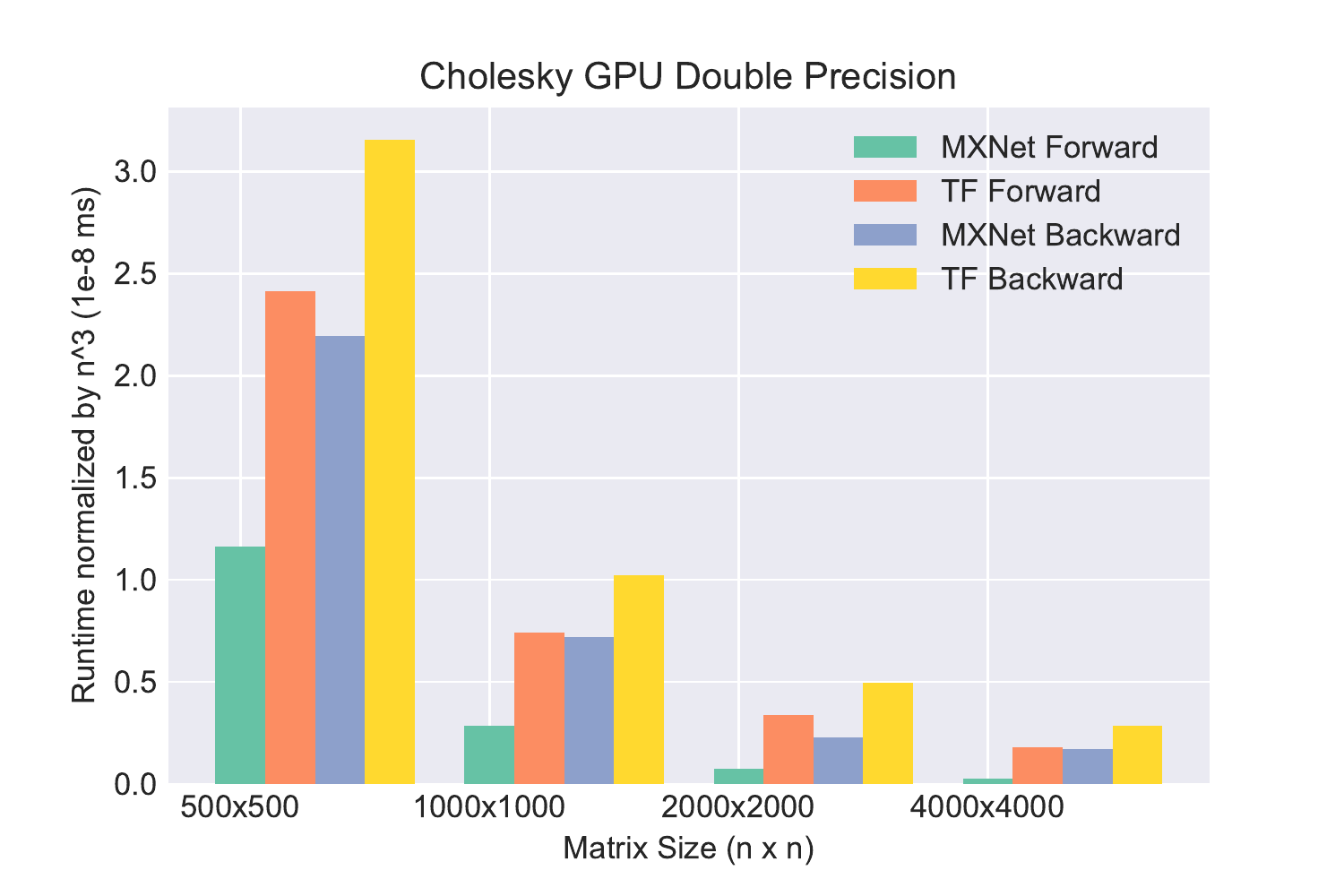}
\endminipage \\
\minipage{0.49\textwidth}
  \includegraphics[width=\linewidth]{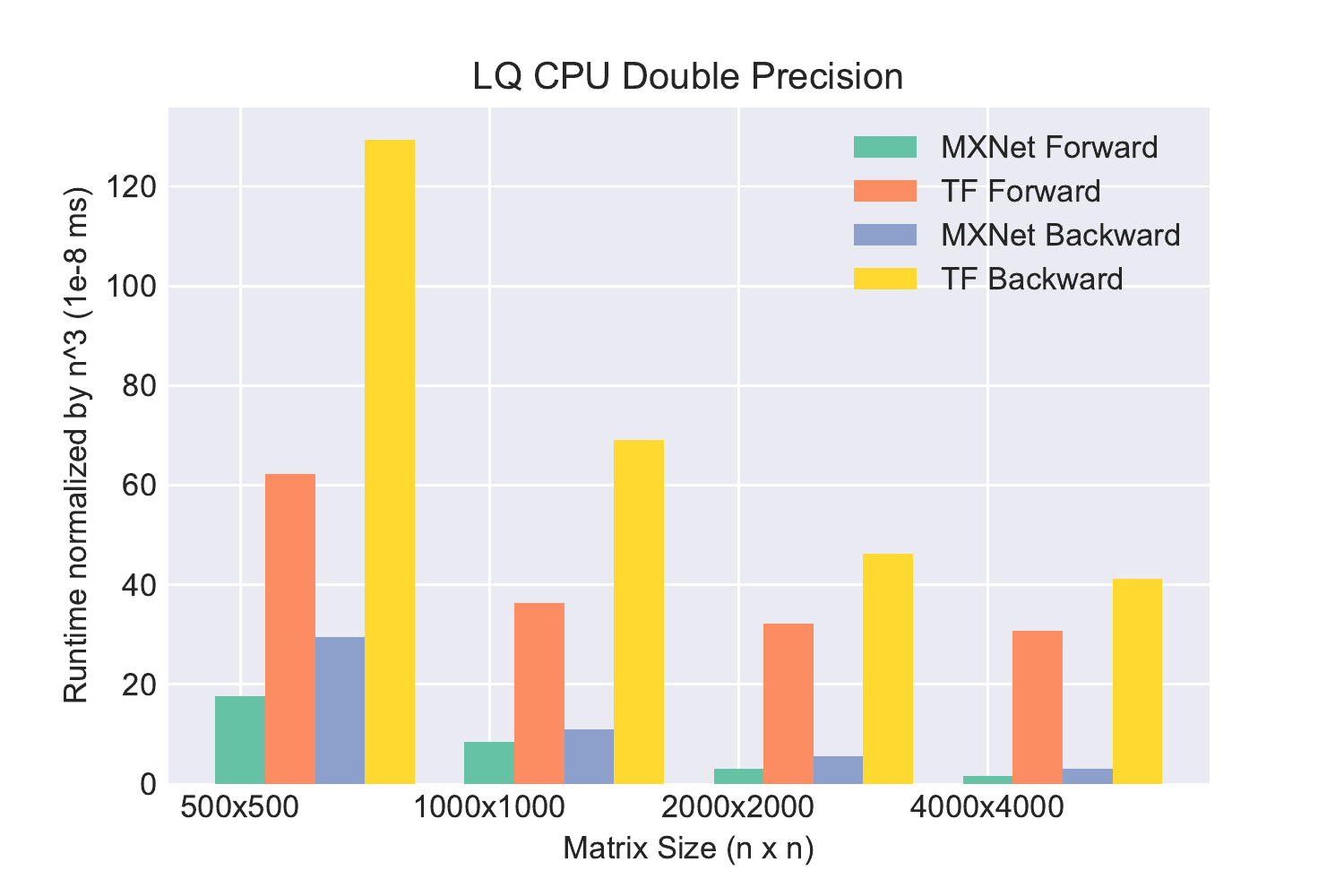}
\endminipage\hfill
\minipage{0.49\textwidth}
  \includegraphics[width=\linewidth]{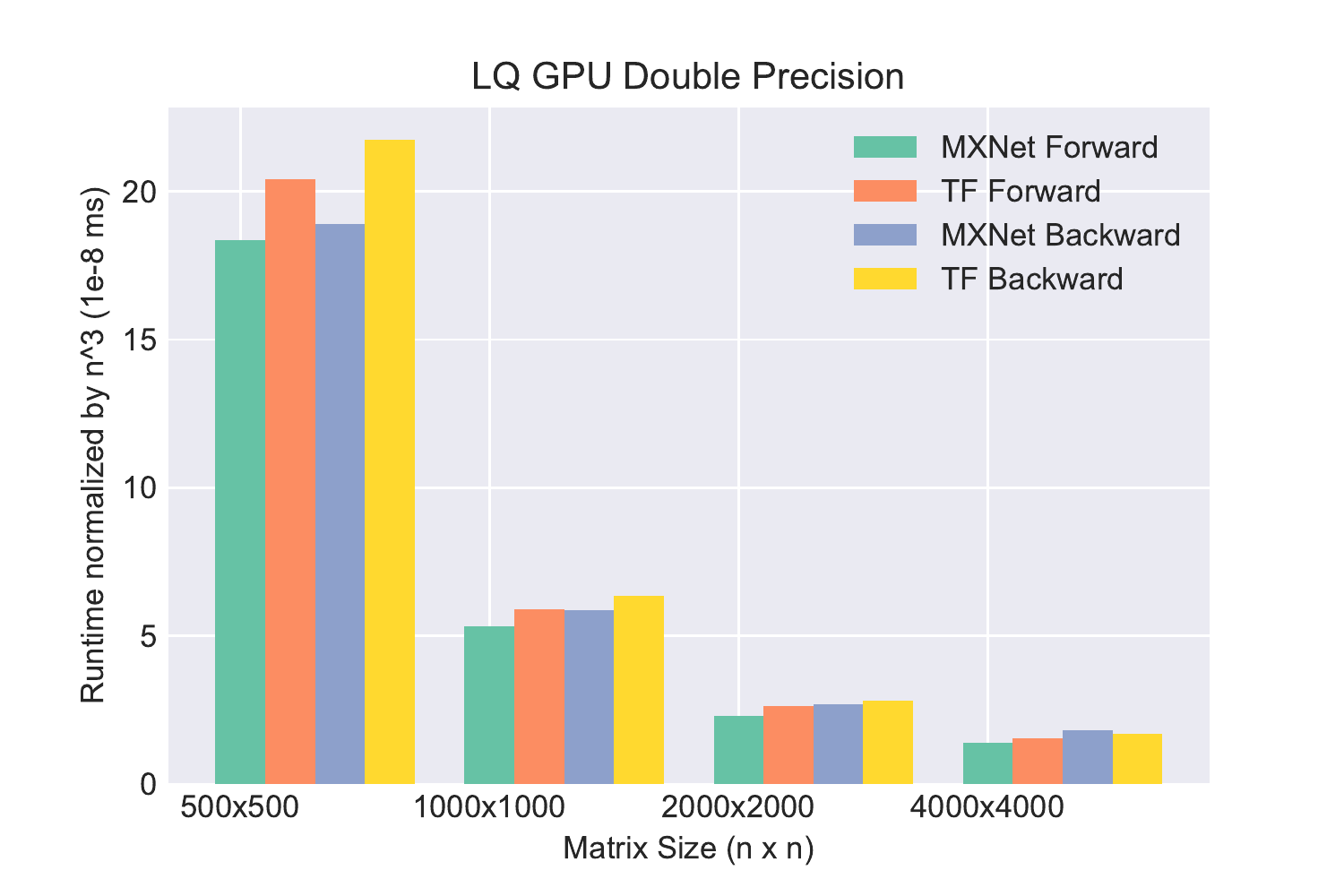}
\endminipage \\
\minipage{0.49\textwidth}
  \includegraphics[width=\linewidth]{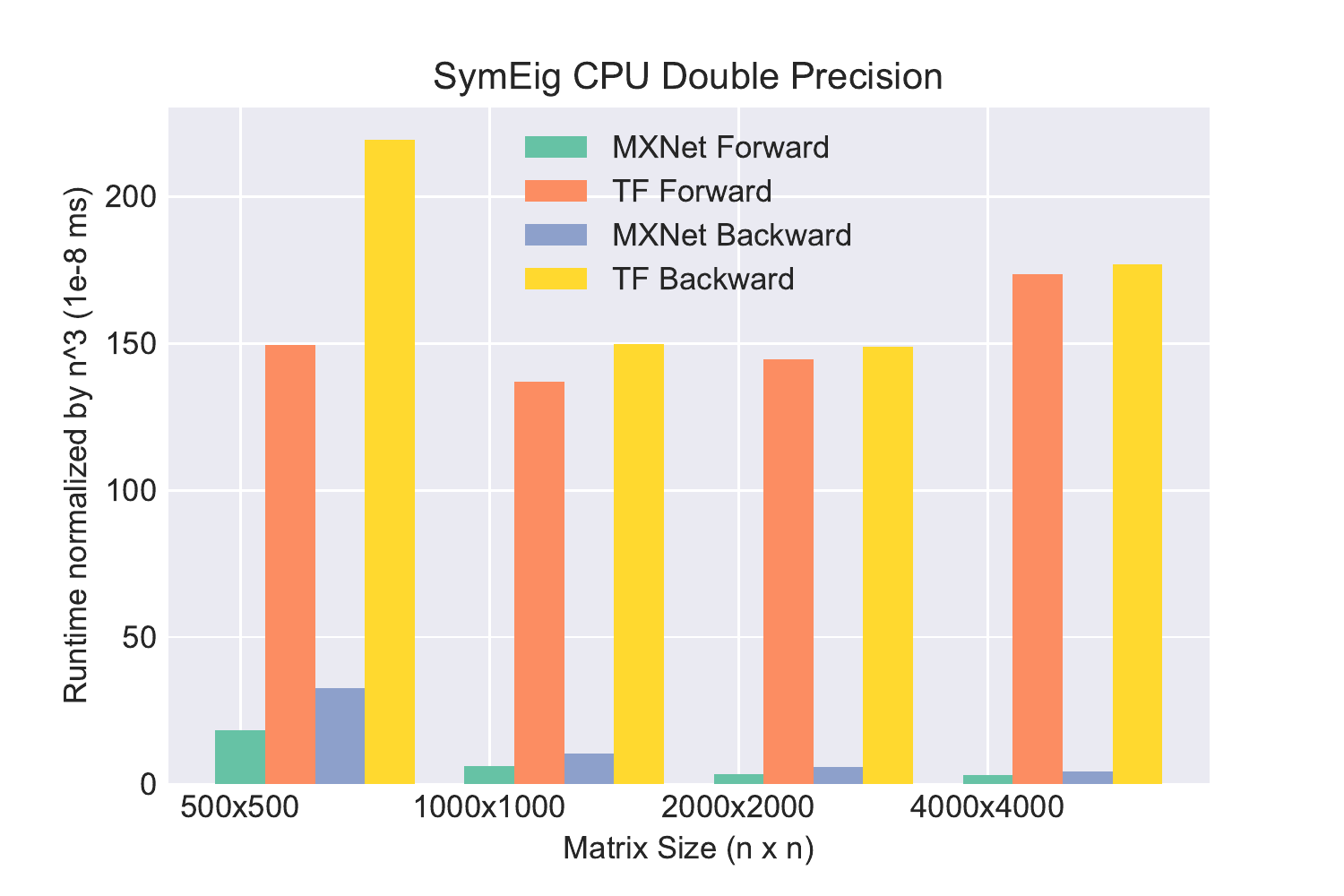}
\endminipage\hfill
\minipage{0.49\textwidth}
  \includegraphics[width=\linewidth]{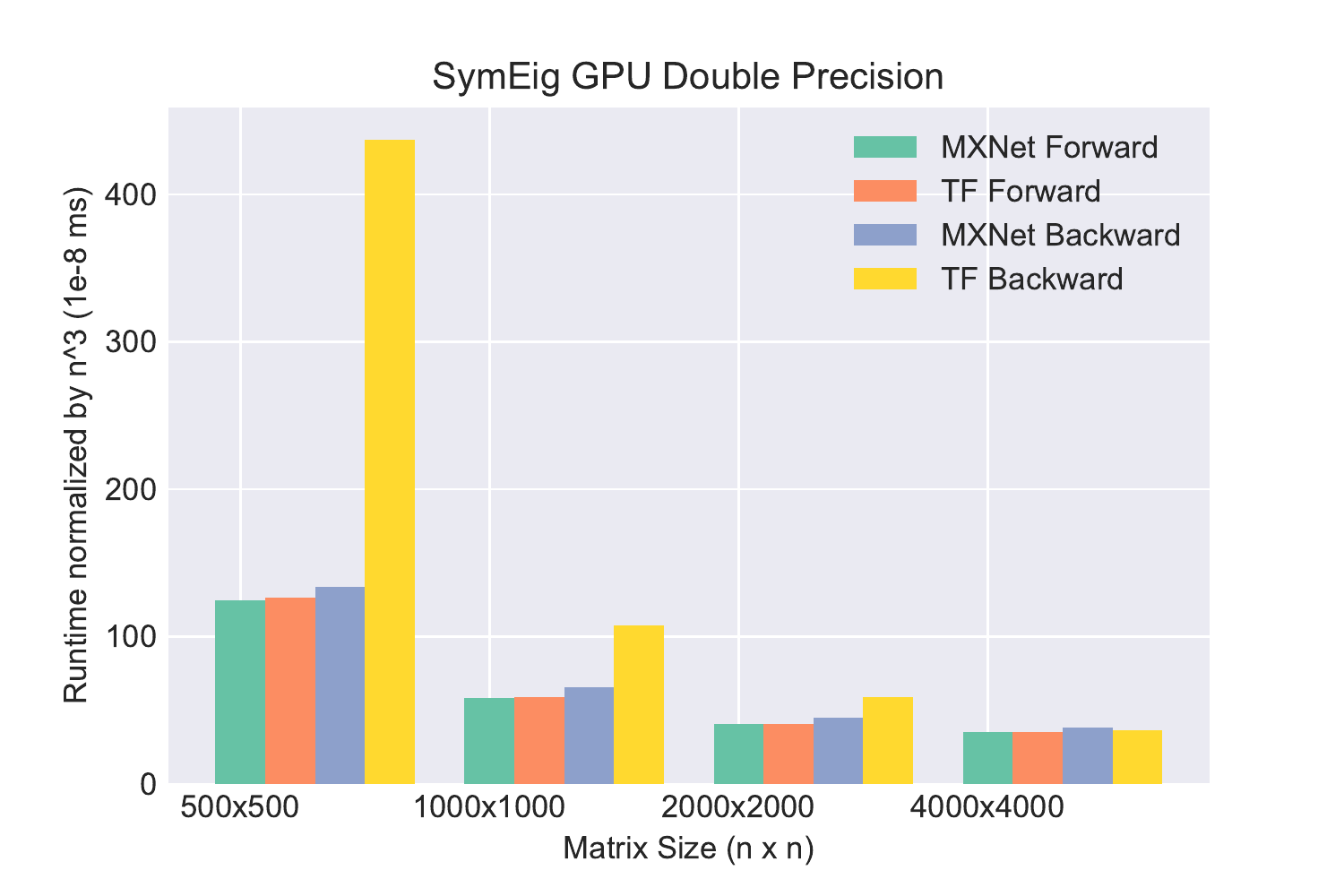}
\endminipage
\caption{\label{fig:potrf-runtime}
  Runtime of $n\times n$ matrix transformations (left column: CPU;
  right column: GPU; top row: Cholesky decomposition; middle row: LQ
  decomposition; bottom row: Symmetric eigen-decomposition). We
  compare our {\tt linalg} extension of MXNet versus TensorFlow, both
  for forward and backward computations. Results are averaged over ten
  repetitions, then normalized by $n^3$.}
\end{figure*}

Results of a runtime comparison for a number of operators are shown in \figref{potrf-runtime}. On the CPU, our code consistently runs more than three times faster. On the GPU, our code generally runs faster than TensorFlow, but the speedup is less pronounced. These differences may be due to us calling optimized BLAS/LAPACK functions directly, taking care to avoid unnecessary allocation of temporary space. Also note that forward runs substantially faster than backward, which is likely due to the composite backward expressions. As noted in \cite{Murray:16}, a direct implementation of backward may offer improvements here, yet would require a substantially more complex implementation.

\subsection{Sparse Gaussian Process}
\label{sec:exper-sgp}

As detailed in \secref{ex-spgp}, we implemented the sparse Gaussian process approximation of \citet{Titsias:09} in MXNet, using our {\tt linalg} extension. Code snippets above indicate the major reduction in implementational complexity, compared to the reference implementation in GPy. Here, we compare the runtime for criterion and gradient computation between these two codes.

\begin{figure}[h!]
\centering
\includegraphics[width=0.5\textwidth]{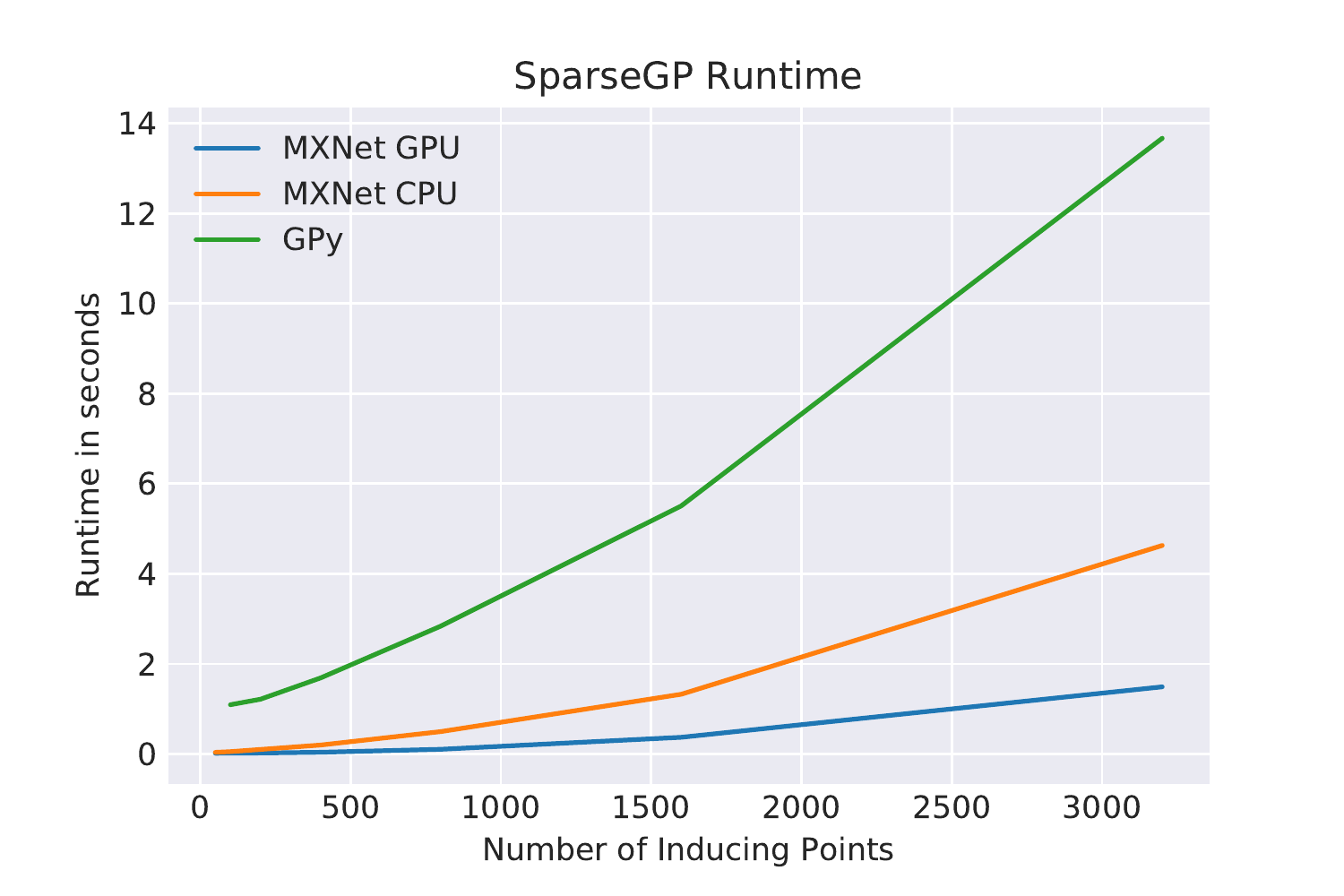}
\caption{\label{fig:sgp-runtime}
  Runtime of variational sparse GP criterion and gradient computation (see
  \secref{ex-spgp}) on the "power" dataset, using our MXNet {\tt linalg} implementation (on CPU and GPU,
  averaged over 100 runs) and the GPy reference implementation (same CPU
  instance, averaged over 20 runs). The x-axis shows the number $U$ of inducing
  points. For $U=50$, MXNet with GPU takes 0.015 seconds, MXNet CPU 0.035 seconds, and GPy 1.070 seconds. For $U=3200$, MXNet GPU takes 1.49 seconds, MXNet CPU 4.63 seconds, and GPy 13.67 seconds.}
\end{figure}

We take a standard regression benchmarking dataset about electrical energy output of a combined cycle power plant \citep{Kaya:12}, called {\em power} below. It has $n=9568$ cases of input dimension four. We measure the wall-clock time of the computation of variational sparse GP criterion and gradient, which the real-world time spent to compute a gradient step of sparse GP optimization. The wall-clock time is measured as the time spent from the starting of the function call and the end of the function call, which is averaged over 20 repetitive runs. We compare the resulting wall-clock time of our MXNet implementation (CPU and GPU) with the GPy implementation (CPU only). Results are provided in \figref{sgp-runtime}. For a large number $U$ of inducing points, our MXNet {\tt linalg} implementation is three times faster than GPy on CPU, and ten times faster than GPy when running on a GPU. When $U$ is small, GPy has a high amount of overhead, this results into our MXNet implementation being up to 70 times faster.

We also compared the memory consumption of GPy vs MXNet, by roughly measuring the peak memory consumption during optimization. We again run MXNet and GPy on CPU with the {\em power} dataset. Throughout optimization, GPy on average uses 2.8 GB of memory, while MXNet uses only 310 MB. This difference is likely due to a caching mechanism for kernel matrices, implemented in GPy, which at least in this particular example proved to be less efficient than MXNet's highly tuned memory management.

\textbf{Performance Comparisons.} We apply sparse GP implementation of ours versus GPy to four standard regression benchmarks. {\em Naval1} and {\em Naval2} are the same regression dataset with two different regression targets, they have $n=11934$ data points and 16 dimensional inputs. {\em Kin8nm} has $n=8192$ data points and 8 dimensional inputs, and {\em power} was already used above. We used 90\% data points for training and 10\% for testing. All the datasets are normalized column-wise. We used $U=50$ inducing points for GPy, and $U=\{50, 3200\}$ inducing points for our MXNet implementation. The Adam optimizer \cite{Kingma:15} was used for training in MXNet, running 3000 iterations with step rate $10^{-2}$, while GPy models are trained for a maximum number of 1000 iterations with L-BFGS. Results with 10-fold averages measured in terms of root mean squared error (RMSE) and test set log likelihood (TLL) are provided in \tabref{sgp-results}. GPy and MXNet with 50 inducing points give similar performance for both RMSE and TLL. By running on GPU, the MXNet implementation allows us to use 3200 inducing points. This gives significantly better performance on all the datasets for both RMSE and TLL.

\begin{table*}[h]
\centering
\begin{tabular}{|l|l|l|l|l|l|}
\hline
Metric               & Method    & naval1 & naval2 & kin8nm & power \\ \hline
\multirow{6}{*}{RMSE} & GPy (50) & $\num{3.5e-5}  \pm$      &    $\num{3.1e-4}  \pm$    &  $\num{8.7e-2}  \pm$      &   $\num{3.98} 	  \pm$    \\
&  & $\num{1.0e-5}$      &    $\num{9.0e-6}$    &  $\num{3.1e-3}$      &   $\num{1.86e-1}$    \\ \cline{2-6}
                     & MXNet (50)   & $\num{3.7e-5}  \pm$      &    $\num{3.1e-4} \pm$    &  $\num{8.85e-2}  \pm$      &   $\num{3.98}  \pm$    \\
                     & & $\num{1.1e-5}$      &    $\num{8.0e-6}$    &  $\num{2.96e-3}$      &   $\num{1.86e-1}$    \\ \cline{2-6}
                     & MXNet (3200) & $\num{0.6e-5}  \pm$     &    $\num{2.97e-4} \pm$    &  $\num{6.79e-2}  \pm$      &   $\num{3.08} \pm$     \\
                     &  & $\num{0.6e-5}$     &    $\num{5.0e-6}$    &  $\num{2.48e-3}$      &   $\num{2.80e-1}$     \\ \hline
\multirow{3}{*}{TLL} & GPy (50)       & $\num{8.58}  \pm  	\num{0.22}$      &    $\num{6.66}  \pm  	\num{0.03}$    &  $\num{0.98}  \pm \num{0.02}$      &   $-\num{2.80}  \pm \num{0.05}$     \\ \cline{2-6}
                     & MXNet (50)   & $\num{8.69}  \pm \num{0.19}$      &    $\num{6.67} 	  \pm \num{0.02}$    &  $\num{0.98}  \pm \num{0.02}$      &   $-\num{2.80}  \pm \num{ 0.05}$     \\ \cline{2-6}
                     & MXNet (3200) & $\num{10.70} 	  \pm \num{0.51}$      &    $\num{6.70}  \pm \num{0.02}$    &  $\num{1.28}  \pm \num{0.04}$      &   $-\num{2.53}  \pm \num{0.10}$     \\ \hline
\end{tabular}
\caption{\label{tab:sgp-results} Performance comparison of sparse GP with MXNet and GPy implementation on four standard regression benchmarks. We compare GPy with 50 inducing points with MXNet with 50 and 3200 inducing points.}
\end{table*}

\subsection{Deep Gaussian Processes}
\label{sec:exper-deepgp}

The benefits of specifying GP models in MXNet, enhanced with our {\tt linalg} operators, increase with model complexity. Not only do all gradients come for free, the derivation and implementation of which usually dominate even first prototyping efforts. Also, since multi-threaded CPU and GPU are supported for the same model specification, even complex models can be run on large datasets with little extra effort. In this section, we focus on Deep Gaussian Processes (DGP) \citep{Damianou:13}, a model of considerable complexity. Unsupervised DGPs are layered stacks of GP models, for example:
\[
  P(\vy{}) = \int P(\vy{} | \vh{1}) P(\vh{1} | \vh{2}) P(\vh{2})\, d\vh{1} d\vh{2}.
\]
Here, both $P(\vy{} | \vh{1})$ and $P(\vh{1} | \vh{2})$ are Gaussian processes, while $P(\vh{2})$ is a white Gaussian. We would like to minimize the criterion $-\log P(\vy{})$ w.r.t.\ parameters, which is hard both due to the intractable integrals and the non-parametric nature of the GP integrands. In DGP, we minimize a variational upper bound instead, which is derived by first applying a standard (structured mean field) bound, then as a second step techniques introduced by \citet{Titsias:09}, see also \secref{ex-spgp}.

In our experiments, we focus on a recent combination of DGPs with stacked variational auto-encoders (VAEs) \cite{Kingma:14}, known as VAE-DGP \citep{Dai:16}. Any variational learning method can be decomposed into a generative model (or ``decoder'') and a recognition model (or ``encoder''). The former is the model of interest (observed variables, conditioned on latent ones), while the latter is the variational distribution used in the bound (latent variables, conditioned on observed ones). The key idea of VAE is to parameterize the recognition model by a neural network, mapping to means and variances of Gaussian distributions. In VAE-DGP, this idea is adopted for the recognition model, while the generative model is a DGP. Learning free parameters in VAE-DGP is of considerable complexity: we have to jointly update generative parameters (i.e., GP hyper-parameters and inducing points) and recognition parameters (DNN weights and biases), computing gradients of a criterion which depends both on sparse GP and deep NN models. Given these complexities, the implementation we benchmark here is remarkably succinct. We started with code for a normal stacked VAE, readily available in MXNet, and replaced terms of the generative model in each layer by the bound of \citet{Titsias:09}, using the code given in \secref{ex-spgp}. Despite its implementational simplicity, this model improves upon VAE-DGP \citep{Dai:16}, both by allowing mini-batch learning and by modeling correlations of latent variables across layers in the variational posterior. It is de facto a new model, termed {\em Stochastic Variational Deep Gaussian Processes} (SV-DGP).

We use an unsupervised learning problem on the MNIST dataset in order to compare the performance of SV-DGP (our MXNet-{\tt linalg} implementation) with results from VAE-DGP (GPy implementation), as well as other previous work. The experimental setup is detailed in \citep[Sect.~5.1]{Dai:16}. The evaluation metric is log-likelihood of test points, which we estimate by importance sampling (using 1000 samples). We evaluate SV-DGP with three different architectures: 5 (1 hidden layer), 10-50 (2 hidden layers), 5-20-50 (3 hidden layers). Results are given in \tabref{mnist}, comparing SV-DGP not only to VAE-DGP, but also the other baselines used in \cite{Dai:16}. With the same setting, SV-DGP consistently outperforms VAE-DGP.

\begin{table}
\centering
\begin{tabular}{c|c}
\hline
Model & MNIST\\\hline\hline
Deep GSN & 214 $\pm$ 1.1 \\
GAN & 225 $\pm$ 2 \\
GMMN+AE & 282 $\pm$ 2\\
\hline
VAE-DGP (5) & 301.67 \\
VAE-DGP (10-50) & 674.86 \\
VAE-DGP (5-20-50) & 723.65\\
\hline
SV-DGP (5) & 313.07\\
SV-DGP (10-50) & 705.03 \\
SV-DGP (5-20-50) & 724.83 \\
\hline
\end{tabular}
\caption{Log-likelihood on MNIST test data. The baselines are VAE-DGP \citep{Dai:16}, Deep GSN \citep{Bengio:14}, GAN \citep{Goodfellow:14} and GMMN+AE \citep{Li:15}.}\label{tab:mnist}
\end{table}

\section{Discussion}
\label{sec:discuss}

We have shown how to introduce advanced linear algebra routines, such as Cholesky factorization, backsubstitution, LQ factorization, and symmetric eigen decomposition, as differentiable operators into a deep learning development system (DLDS). Implemented in a time and memory efficient manner, they open the door to realizing a host of ``pre-deep'' machine learning models within a DLDS, and enable combinations of Gaussian processes, Bayesian linear models, or linear dynamical systems with deep multi-layer maps. We provide a range of machine learning examples, complete notebooks for which are available for download.

All operators mentioned here are implemented in MXNet (\url{https://github.com/apache/incubator-mxnet}). In contrast to similar operators in TensorFlow and Theano, our implementation is tuned for maximum memory efficiency. We call highly optimized BLAS and LAPACK library functions directly on the C++ level. In particular, we take care to use the minimum required amount of temporary storage. Most of our operators are implemented in-place, meaning that no additional memory is needed beyond inputs and outputs. If memory-intensive machine learning or scientific computing methods (such as Gaussian processes, least squares estimation, Kalman smoothing, or principal components analysis) are to be powered by easy-to-use DLDS, it will be essential to implement key linear algebra operators with the same level of care that is applied to convolution or LSTM layers today.

\subsubsection*{Acknowledgments}

We would like to thank Yuyang (Bernie) Wang, Jan Gasthaus and Syama Rangapuram for providing the ``Time Series'' chapter in the ``Deep Learning: The Straight Dope'' online book, containing a full-fledged Kalman filtering implementation, using our novel operators. Moreover, we would like to thank the core developers and the open source community around MXNet for much help and support.


\bibliography{../../template/papers,../../template/books,../icml2018/autodiff}


\section*{Appendix}

In this section, we collect derivations for {\em backward} expressions of supported operators. Note that some operators are not covered, in case the respective derivations are obvious, given the ones reported here.

\subsubsection*{potrf}

This derivation is given in \cite{Murray:16}.
Recall that $\mxa{} = \mxl{}\mxl{}^T$, where $\mxa{}$ is symmetric, $\mxl{}$ is lower triangular. We first determine $d\mxa{}$ in terms of $d\mxl{}$ and inputs. Differentiate the defining equation, and apply $\mxl{}^{-1}(\cdot)\mxl{}^{-T}$:
\[
  \mxl{}^{-1}(d\mxa{})\mxl{}^{-T} = 2\sym\left( \mxl{}^{-1}(d\mxl{}) \right),
\]
Importantly, $\mxl{}^{-1}(d\mxl{})$ is lower triangular, so we really only have to equate the lower triangle on both sides (taking care of the diagonal). Define the lower triangular ``masking matrix'' $\mxe{}\in\R^{n\times n}$ by
\[
  e_{i j} = \left\{ \begin{array}{ll}
    0 & |\;\; i < j \\
    1 & |\;\; i = j \\
    2 & |\;\; i > j
  \end{array}\right\}.
\]
Then we have:
\[
   d\mxl{} = \frac{1}2 \mxl{}\left( \mxl{}^{-1}(d\mxa{})\mxl{}^{-T} \circ \mxe{}
  \right).
\]
For the {\em backward} expression:
\[
  d\phi = \trace\bmxl{}{}^T(d\mxl{}) = \frac{1}2 \trace\bmxl{}{}^T\mxl{}(
  \mxl{}^{-1}(d\mxa{})\mxl{}^{-T} \circ \mxe{} ) = \frac{1}2\trace(
  \mxl{}^T\bmxl{} \circ \mxe{} )^T \mxl{}^{-1}(d\mxa{})\mxl{}^{-T}.
\]
We used that $\trace\mxx{}^T(\mxy{}\circ\mxe{}) = \trace(\mxx{}\circ\mxe{})^T\mxy{}$. Note that $\bmxa{}$ must be symmetric. Therefore:
\[
  \bmxa{} = \frac{1}2 \mxl{}^{-T} \sym(\mxl{}^T\bmxl{} \circ \mxe{}) \mxl{}^{-1}.
\]
Note that $\sym(\mxx{}\circ\mxe{}) = \text{copyltu}(\mxx{})$, so that
\[
  \bmxa{} = \frac{1}2 \mxl{}^{-T} \text{copyltu}(\mxl{}^T\bmxl{}) \mxl{}^{-1}.
\]

If the Cholesky factor is upper triangular, in that $\mxa{} = \mxr{}^T\mxr{}$, where $\mxr{}$ is upper triangular, we can use the derivation above, and then $\mxr{} = \mxl{}^T$. Therefore,
\[
  \bmxa{} = \frac{1}2 \mxr{}^{-1} \text{copyltu}(\mxr{}\bmxr{}{}^T) \mxr{}^{-T} =
   \frac{1}2 \mxr{}^{-1} \text{copyutl}(\bmxr{}\mxr{}^T) \mxr{}^{-T}.
\]
Here $\text{copyutl}(\mxx{})$ creates a symmetric matrix by copying the upper triangle of $\mxx{}$ to its lower triangle.

\subsubsection*{gelqf}

To the best of our knowledge, this derivation is novel.
Recall that $\mxa{} = \mxl{}\mxq{}$, where $\mxa{}, \mxq{}\in\R^{m\times n}$, $\mxl{}\in\R^{m\times m}$. Also, $\mxq{}\mxq{}^T = \Id_m$, and $\mxl{}$ is lower triangular with nonnegative diagonal. First:
\[
  d\mxa{} = (d\mxl{})\mxq{} + \mxl{}(d\mxq{})\quad\Rightarrow\quad
  d\mxq{} = \mxl{}^{-1}\left( d\mxa{} - (d\mxl{})\mxq{} \right).
\]
Differentiating $\mxq{}\mxq{}^T = \Id_m$, we note that $(d\mxq{})\mxq{}^T$ is skew-symmetric (meaning that $\mxx{}^T = -\mxx{}$). Multiplying the above equation by $\mxq{}^T$ on the right:
\[
  (d\mxq{})\mxq{}^T = \mxc{} - \mxl{}^{-1}(d\mxl{}),\quad \mxc{} := \mxl{}^{-1}
  (d\mxa{})\mxq{}^T.
\]
Using the skew-symmetry:
\[
  \mxc{} - \mxl{}^{-1}(d\mxl{}) = ( \mxl{}^{-1}(d\mxl{}) )^T - \mxc{}^T\quad
  \Rightarrow\quad
  \sym(\mxc{}) = \sym(\mxl{}^{-1}(d\mxl{})).
\]
Since $\mxl{}^{-1}(d\mxl{})$ is lower triangular, we use the same argumentation as for {\tt potrf} above:
\[
  d\mxl{} = \mxl{}\left( \sym(\mxc{}) \circ \mxe{} \right).
\]
Plugging these into $d\phi$, we obtain
\[
  d\phi = \trace\left( \mxl{}^{-T}\bmxq{} \right)^T (d\mxa{}) + \trace\left(
  \bmxl{} - \mxl{}^{-T} \bmxq{}\mxq{}^T \right)^T (d\mxl{}).
\]
Plugging $d\mxl{}$ into the second term:
\[
  \trace(\dots)^T (d\mxl{}) = \trace\mxm{}^T (\sym(\mxc{})\circ\mxe{}) =
  \trace\sym(\mxm{}\circ\mxe{}) \mxc{},\quad \mxm{} := \mxl{}^T\bmxl{} -
  \bmxq{}\mxq{}^T.
\]
Plugging in $\mxc{}$, some further algebra gives
\[
  \bmxa{} = \mxl{}^{-T}\left( \bmxq{} + \sym(\mxm{}\circ\mxe{}) \mxq{}
  \right) = \mxl{}^{-T}\left( \bmxq{} + \text{copyltu}(\mxm{}) \mxq{} \right).
\]

\subsubsection*{syevd}

This derivation is given in \cite{Giles:08}.
Let $\mxa{}\in\R^{n\times n}$ be symmetric. Recall that $\mxu{}\mxa{} = \mxlam{}\mxu{}$, where $\mxlam{}=\diag\vlam{}$, and $\mxu{}\mxu{}^T = \Id$. Differentiating the equation and right-multiplication with $\mxu{}^T$ gives:
\[
  \mxu{}(d\mxa{})\mxu{}^T + (d\mxu{})\mxu{}^T\mxlam{} = \mxlam{}
  (d\mxu{})\mxu{}^T + d\mxlam{},
\]
where we used that $\mxa{}\mxu{}^T = \mxu{}^T\mxlam{}$. Denote
\[
  \mxm{} := \mxu{}(d\mxa{})\mxu{}^T,\quad \mxs{} := (d\mxu{})\mxu{}^T.
\]
Differentiating $\mxu{}\mxu{}^T = \Id$, we see that $\mxs{}$ is skew-symmetric, and in particular $\diag(\mxs{}) = \vzero$. The equation becomes
\[
  \mxm{} + \mxs{}\mxlam{} = \mxlam{}\mxs{} + d\mxlam{}.
\]
Now, $\diag(\mxlam{}\mxs{}) = \vlam{}\circ\diag(\mxs{}) = \vzero$, so that
\[
  d\vlam{} = \diag(\mxm{})
\]
Denote $\tmxm{} := \mxm{} - (\mxm{}\circ\Id)$, whose diagonal is zero. Now:
\[
  \mxlam{}\mxs{} - \mxs{}\mxlam{} = \left[ s_{i j}(\lambda_i-\lambda_j)
  \right]_{i,j} = \tmxm{}.
\]
Therefore: $s_{i j} = \tsm{i j}/(\lambda_i-\lambda_j)$ for $i\ne j$, and $s_{i i} = 0$. Define the matrix $\mxf{}\in\R^{n\times n}$ as
\[
  F_{i j} = \frac{\Ind{i\ne j}}{\lambda_i - \lambda_j},
\]
where the indicator has preference, so that $F_{i i} = 0$. Then: $\mxs{} = \tmxm{}\circ\mxf{}$, therefore
\[
  d\mxu{} = (\tmxm{}\circ\mxf{}) \mxu{} = (\mxm{}\circ\mxf{}) \mxu{},
\]
since $\diag(\mxf{}) = \vzero$. Then:
\[
  d\phi = \trace\bmxu{}{}^T(d\mxu{}) + \bvlam{}{}^T(d\vlam{}) =
  \trace\mxu{}\bmxu{}{}^T (\mxm{}\circ\mxf{}) + \trace\bmxlam{}\mxm{} =
  \trace\left( \bmxu{}\mxu{}^T \circ \mxf{} + \bmxlam{} \right)^T \mxm{}.
\]
Plugging in $\mxm{}$, and noting that it is symmetric, some algebra gives:
\[
  \bmxa{} = \mxu{}^T\left( \sym(\bmxu{}\mxu{}^T \circ \mxf{}) + \bmxlam{}
  \right) \mxu{}.
\]
When working with $\mxf{}$, we need to guard against division by zero. Therefore, we use a slight modification:
\[
  F_{i j} = \frac{\Ind{i\ne j}}{h(\lambda_i - \lambda_j)},\quad
  h(t) = \max(|t|,\eps) \sgn(t),
\]
and $\eps>0$ is a small scalar.

\ifsvd

\subsubsection*{Singular Value Decomposition (gesvd)}

Let $\mxa{}\in\R^{m\times n}$, where $m\le n$. The singular value decomposition (SVD) is given by
\[
  \mxa{} = \mxu{}^T\mxlam{}\mxv{},\quad \mxlam{} = \diag\vlam{}\in
  \R^{m\times m},\; \mxu{}\in\R^{m\times m},\; \mxv{}\in\R^{m\times n}.
\]
Here, $\mxu{}$ and $\mxv{}$ are orthonormal:
\[
  \mxu{}\mxu{}^T = \mxu{}^T\mxu{} = \Id_m,\quad \mxv{}\mxv{}^T = \Id_m.
\]
The singular values $\lambda_i$ are non-zero and ascending:
\[
  0\le \lambda_1\le\lambda_2\le\dots\le\lambda_m.
\]
Other ways to define the SVD are
\begin{equation}\label{eq:svd-def}
  \mxu{}\mxa{} = \mxlam{}\mxv{},\quad \mxa{}\mxv{}^T = \mxu{}^T\mxlam{}.
\end{equation}
The {\tt backward} expressions are well-defined under the following extra conditions that the singular values are distinct and positive:
\[
  0 < \lambda_1 < \lambda_2 < \dots < \lambda_m.
\]
In particular, the matrix $\mxa{}$ must have full rank $m$. Our implementation does not fail if some singular values are equal, but the backward expression is wrong then (the exact expression being undefined).

We first need to derive expressions for $d\vlam{}$, $d\mxu{}$, $d\mxv{}$. Differentiating \eqp{svd-def} and right-multiplication by $\mxv{}^T$:
\[
  \mxu{}(d\mxa{})\mxv{}^T + (d\mxu{})\mxa{}\mxv{}^T = d\mxlam{} +
  \mxlam{}(d\mxv{})\mxv{}^T.
\]
Define
\[
  \mxm{} := \mxu{}(d\mxa{})\mxv{}^T,\quad \mxs{u} := (d\mxu{})\mxu{}^T,\quad
  \mxs{v} := (d\mxv{})\mxv{}^T.
\]
Using the right side of \eqp{svd-def}, we have:
\[
  \mxm{} + \mxs{u}\mxlam{} = d\mxlam{} + \mxlam{}\mxs{v}.
\]
Here, $\mxs{u}, \mxs{v}$ are skew-symmetric:
\[
  \mxs{*}^T = -\mxs{*},\quad \diag(\mxs{*}) = \vzero,\quad * \in \{u, v\}.
\]
Therefore:
\[
  d\vlam{} = \diag(\mxm{}).
\]
Using the notation $\tmxm{} = \mxm{} - (\mxm{}\circ\Id)$ (so $\mxm{}$ with its diagonal set to zero), we have that
\[
   \tmxm{} = \mxlam{}\mxs{v} - \mxs{u}\mxlam{}.
\]
Using the skew-symmetry:
\[
  \tmxm{}{}^T = \mxlam{}\mxs{u} - \mxs{v}\mxlam{}.
\]
Define
\[
  \mxm{*} := \frac{1}2\left( \mxm{} * \mxm{}^T \right),\quad \mxs{*} :=
  \frac{1}2\left( \mxs{v} * \mxs{u} \right),\quad *\in \{+, -\}.
\]
Then:
\[
  \tmxm{+} = \mxlam{}\mxs{+} - \mxs{+}\mxlam{},\quad \tmxm{-} = \mxlam{}
  \mxs{-} + \mxs{-}\mxlam{}.
\]
Define
\[
  \mxf{*} = \left[ \frac{\Ind{i\ne j}}{\lambda_i * \lambda_j} \right]_{i,j},\quad *\in
  \{+, -\}.
\]
Also, define
\[
  \mxe{} := \mxf{+}\circ\mxf{-} = \left[
  \frac{\Ind{i\ne j}}{\lambda_i^2 - \lambda_j^2} \right]_{i,j}.
\]
As for {\tt syevd}, we have
\[
  \mxs{+} = \tmxm{+} \circ \mxf{-} = \mxm{+}\circ\mxf{-},
\]
because $\diag(\mxf{-}) = \vzero$, and
\[
  \mxs{-} = \tmxm{-} \circ \mxf{+} = \mxm{-}\circ\mxf{+}.
\]
Finally,
\[
  \mxs{u} = (d\mxu{})\mxu{}^T = \mxs{+} - \mxs{-} = \mxm{+}\circ\mxf{-} -
  \mxm{-}\circ\mxf{+} = \mxm{}\circ \frac{1}2(\mxf{-} - \mxf{+}) + \mxm{}^T
  \circ \frac{1}2(\mxf{-} + \mxf{+}).
\]
We can also get an expression for $\mxs{v}$, but this is not needed in what follows, because $d\mxv{}$ cannot be obtained from $\mxs{v}$. For $i\ne j$:
\[
  \frac{1}{\lambda_i - \lambda_j} - \frac{1}{\lambda_i + \lambda_j} =
  \frac{2\lambda_j}{\lambda_i^2 - \lambda_j^2}.
\]
Some algebra gives:
\[
  \frac{1}2(\mxf{-} - \mxf{+}) = \mxe{}\mxlam{},\quad
  \frac{1}2(\mxf{-} + \mxf{+}) = \mxlam{}\mxe{},
\]
therefore
\[
  \mxs{u} = \mxm{}\circ(\mxe{}\mxlam{}) + \mxm{}^T\circ(\mxlam{}\mxe{}) =
  \left(\mxm{}\mxlam{} + \mxlam{}\mxm{}^T \right)\circ\mxe{} =
  2\sym(\mxm{}\mxlam{})\circ\mxe{}
\]
and
\[
  d\mxu{} = 2\left( \sym(\mxm{}\mxlam{})\circ\mxe{} \right) \mxu{}.
\]
An expression for $d\mxv{}$ cannot be obtained from $\mxs{v}$. We can differentiate \eqp{svd-def}, and use
\[
  (d\mxu{})\mxa{} = (d\mxu{})\mxu{}^T\mxu{}\mxa{} = \mxs{u}\mxlam{}\mxv{}
\]
to obtain
\[
  d\mxv{} = \mxlam{}^{-1}\left( (\mxs{u}\mxlam{} - d\mxlam{})\mxv{} + \mxu{}
  (d\mxa{}) \right).
\]
Here, we require that all $\lambda_i>0$. Next, we have to bring
\[
  d\phi = \bvlam{}{}^T(d\vlam{}) + \trace\bmxu{}{}^T(d\mxu{}) + \trace\bmxv{}{}^T
  (d\mxv{})
\]
into the form $d\phi = \trace\bmxa{}{}^T(d\mxa{})$. Denote the terms by (a), (b), (c).
First,
\[
  (a) = \bvlam{}{}^T(d\vlam{}) = \trace\bmxlam{}{}^T\mxu{}(d\mxa{})\mxv{}^T =
  \trace(\mxu{}^T\bmxlam{}\mxv{})^T(d\mxa{}).
\]
Next,
\[
  (b) = \trace\bmxu{}{}^T(d\mxu{}) = \trace(\bmxu{}\mxu{}^T)^T\mxs{u}.
\]
We come back to this term below. Next,
\[
  (c) = \trace\bmxv{}{}^T\mxlam{}^{-1}\left( (\mxs{u}\mxlam{} - d\mxlam{})\mxv{}
  + \mxu{} (d\mxa{}) \right) = \trace(\mxu{}^T\mxlam{}^{-1}\bmxv{})^T(d\mxa{})
  + \trace\bmxv{}{}^T\mxlam{}^{-1}(\mxs{u}\mxlam{} - d\mxlam{})\mxv{}.
\]
Denote the final trace term by $(d) + (e)$. First,
\[
\begin{split}
  (e) & = -\trace\bmxv{}{}^T\mxlam{}^{-1}(d\mxlam{})\mxv{} = -\trace(\mxlam{}^{-1}
  \bmxv{}\mxv{}^T)^T(d\mxlam{}) = -\trace(\mxlam{}^{-1}\bmxv{}\mxv{}^T \circ\Id)
  \mxu{}(d\mxa{})\mxv{}^T \\
  & = -\trace\left( \mxu{}^T(\mxlam{}^{-1}\bmxv{}\mxv{}^T
  \circ\Id)\mxv{} \right)^T(d\mxa{}).
\end{split}
\]
Finally,
\[
\begin{split}
  (d) & = \trace\bmxv{}{}^T\mxlam{}^{-1}\mxs{u}\mxlam{}\mxv{} = \trace \mxlam{}
  \mxv{}\bmxv{}{}^T\mxlam{}^{-1}\mxs{u}, \\
  (b) + (d) & = \trace\mxg{1}^T\mxs{u},\quad \mxg{1} := \bmxu{}\mxu{}^T +
  \mxlam{}^{-1}\bmxv{}\mxv{}^T\mxlam{}.
\end{split}
\]
Moreover,
\[
\begin{split}
  \trace\mxg{1}^T\mxs{u} & = 2\trace\mxg{1}^T(\sym(\mxm{}\mxlam{})\circ\mxe{})
  = 2\trace \sym(\mxg{1}\circ\mxe{}) \mxu{}(d\mxa{})\mxv{}^T\mxlam{} \\
  & = \trace\left( 2\mxu{}^T \sym(\mxg{1}\circ\mxe{}) \mxlam{}\mxv{} \right)^T
  (d\mxa{}).
\end{split}
\]
Putting all of this together, we have:
\[
  \bmxa{} = \mxu{}^T\left( \mxg{2}\mxv{} + \mxlam{}^{-1}\bmxv{} \right),\quad
  \mxg{2} = \bmxlam{} + 2\sym(\mxg{1}\circ\mxe{})\mxlam{} - (\mxlam{}^{-1}
  \bmxv{}\mxv{}^T \circ\Id),\quad \mxg{1} = \bmxu{}\mxu{}^T +
  \mxlam{}^{-1}\bmxv{}\mxv{}^T\mxlam{}.
\]
We can compute $\bmxa{}$ with only $O(m^2)$ additional memory:
\begin{itemize}
\item
  $\bmxa{}\leftarrow \mxlam{}^{-1}\bmxv{}$
\item
  $(\mxlam{}^{-1}\bmxv{})\mxv{}^T$, extract diagonal
\item
  $(\mxlam{}^{-1}\bmxv{}\mxv{}^T)\mxlam{}$, then $\mxg{1}$
\item
  $\mxg{2}$, then $\bmxa{}\leftarrow \bmxa{} + \mxg{2}\mxv{}$ ({\tt gemm})
\item
  $\bmxa{}\leftarrow \mxu{}^T\bmxa{}$
\end{itemize}
If the final computation of $\mxu{}^T\bmxa{}$ is done with {\tt gemm}, another $O(n m)$ of memory is needed. This can be avoided by chunking $\bmxa{}$ into $m\times m$ blocks, and calling {\tt gemm} with $\mxu{}^T$ and each of these blocks. Now, the $O(m^2)$ memory used for $\mxg{2}$ is sufficient.

\fi

\subsubsection*{trsm}

We work out {\em backward} for $\mxb{} = \mxl{}^{-1}\mxa{}$, the other cases are similar.
\[
  d\mxb{} = \mxl{}^{-1}(d\mxa{}) - \mxl{}^{-1}(d\mxl{})\mxl{}^{-1}\mxa{}
\]
From this, we obtain
\[
  d\phi = \trace\bmxb{}^T(d\mxb{}) = \trace\left( \mxl{}^{-T}\bmxb{} \right){}^T
  (d\mxa{}) + \trace\left( -\mxl{}^{-T}\bmxb{}\mxa{}^T\mxl{}^{-T} \right)^T
  (d\mxl{}).
\]
Therefore:
\[
  \bmxa{} = \mxl{}^{-T}\bmxb{},\quad \bmxl{} = -\tril\left( \bmxa{}\mxa{}^T
  \mxl{}^{-T} \right) = -\tril\left( \bmxa{}\mxb{}^T \right).
\]
Here, $\tril(\cdot)$ extracts the lower triangle.

Note that $\mxa{}$ is not needed (but $\mxl{}$ is). We have to compute $\bmxa{}$ first. It can overwrite $\bmxb{}$ or $\mxa{}$. Also, $\bmxl{}$ can overwrite $\mxl{}$. The different cases are:
\[
\begin{split}
  \mxb{}=\mxl{}^{-1}\mxa{}\quad & \Rightarrow\quad \bmxa{}=\mxl{}^{-T}\bmxb{},
  \; \bmxl{} = -\tril\left( \bmxa{}\mxb{}^T \right), \\
  \mxb{}=\mxl{}^{-T}\mxa{}\quad & \Rightarrow\quad \bmxa{}=\mxl{}^{-1}\bmxb{},
  \; \bmxl{} = -\tril\left( \mxb{}\bmxa{}{}^T \right), \\
  \mxb{}=\mxa{}\mxl{}^{-1}\quad & \Rightarrow\quad \bmxa{}=\bmxb{}\mxl{}^{-T},
  \; \bmxl{} = -\tril\left( \mxb{}^T\bmxa{} \right), \\
  \mxb{}=\mxa{}\mxl{}^{-T}\quad & \Rightarrow\quad \bmxa{}=\bmxb{}\mxl{}^{-1},
  \; \bmxl{} = -\tril\left( \bmxa{}{}^T\mxb{} \right)
\end{split}
\]

If the triangular matrix $\mxr{}$ is upper triangular instead, we can replace $\mxr{}$ with $\mxl{}^T$ in the derivations. We end up with:
\[
\begin{split}
  \mxb{}=\mxr{}^{-1}\mxa{}\quad & \Rightarrow\quad \bmxa{}=\mxr{}^{-T}\bmxb{},
  \; \bmxr{} = -\triu\left( \bmxa{}\mxb{}^T \right), \\
  \mxb{}=\mxr{}^{-T}\mxa{}\quad & \Rightarrow\quad \bmxa{}=\mxr{}^{-1}\bmxb{},
  \; \bmxr{} = -\triu\left( \mxb{}\bmxa{}{}^T \right), \\
  \mxb{}=\mxa{}\mxr{}^{-1}\quad & \Rightarrow\quad \bmxa{}=\bmxb{}\mxr{}^{-T},
  \; \bmxr{} = -\triu\left( \mxb{}^T\bmxa{} \right), \\
  \mxb{}=\mxa{}\mxr{}^{-T}\quad & \Rightarrow\quad \bmxa{}=\bmxb{}\mxr{}^{-1},
  \; \bmxr{} = -\triu\left( \bmxa{}{}^T\mxb{} \right)
\end{split}
\]
Here, $\triu(\cdot)$ extracts the upper triangle.

\subsubsection*{trmm}

We work out {\em backward} for $\mxb{} = \mxl{}\mxa{}$, the other cases are similar.
\[
  d\mxb{} = (d\mxl{})\mxa{} + \mxl{}(d\mxa{})
\]
gives
\[
  \bmxa{} = \mxl{}^T\bmxb{},\quad \bmxl{} = \tril\left( \bmxb{}\mxa{}^T
  \right).
\]
This can be done in-place. We compute $\bmxl{}$ first, in which case $\bmxa{}$ can overwrite $\bmxb{}$. The different cases are:
\[
\begin{split}
  \mxb{}=\mxl{}\mxa{}\quad & \Rightarrow\quad \bmxa{}=\mxl{}^{T}\bmxb{},
  \; \bmxl{} = \tril\left( \bmxb{}\mxa{}^T \right), \\
  \mxb{}=\mxl{}^T\mxa{}\quad & \Rightarrow\quad \bmxa{}=\mxl{}\bmxb{},
  \; \bmxl{} = \tril\left( \mxa{}\bmxb{}{}^T \right), \\
  \mxb{}=\mxa{}\mxl{}\quad & \Rightarrow\quad \bmxa{}=\bmxb{}\mxl{}^{T},
  \; \bmxl{} = \tril\left( \mxa{}^T\bmxb{} \right), \\
  \mxb{}=\mxa{}\mxl{}^T\quad & \Rightarrow\quad \bmxa{}=\bmxb{}\mxl{},
  \; \bmxl{} = \tril\left( \bmxb{}{}^T\mxa{} \right) \\
\end{split}
\]

If the triangular matrix $\mxr{}$ is upper triangular instead, we can replace $\mxr{}$ with $\mxl{}^T$ in the derivations. We end up with:
\[
\begin{split}
  \mxb{}=\mxr{}\mxa{}\quad & \Rightarrow\quad \bmxa{}=\mxr{}^{T}\bmxb{},
  \; \bmxr{} = \triu\left( \bmxb{}\mxa{}^T \right), \\
  \mxb{}=\mxr{}^T\mxa{}\quad & \Rightarrow\quad \bmxa{}=\mxr{}\bmxb{},
  \; \bmxr{} = \triu\left( \mxa{}\bmxb{}{}^T \right), \\
  \mxb{}=\mxa{}\mxr{}\quad & \Rightarrow\quad \bmxa{}=\bmxb{}\mxr{}^{T},
  \; \bmxr{} = \triu\left( \mxa{}^T\bmxb{} \right), \\
  \mxb{}=\mxa{}\mxr{}^T\quad & \Rightarrow\quad \bmxa{}=\bmxb{}\mxr{},
  \; \bmxr{} = \triu\left( \bmxb{}{}^T\mxa{} \right) \\
\end{split}
\]
Here, $\triu(\cdot)$ extracts the upper triangle.

\subsubsection*{potri}

The operator is $\mxb{} = \mxa{}^{-1} = \texttt{dpotri}(\mxl{})$. Note that $\mxb{}$ is symmetric. We have
\[
  d\mxb{} = -\mxb{}(d\mxa{})\mxb{} = -2\mxb{}\sym\left( (d\mxl{})\mxl{}^T \right)
  \mxb{} = -2\sym\left( \mxb{}(d\mxl{})\mxl{}^{-1} \right),
\]
using $\mxb{} = \mxl{}^{-T}\mxl{}^{-1}$. We must not assume that $\bmxb{}$ is symmetric: $\trace\bmxb{}{}^T\sym(\mxx{}) = \trace\sym(\bmxb{})\mxx{}$. Therefore:
\[
  d\phi = -2 \trace\sym(\bmxb{})\mxb{}(d\mxl{})\mxl{}^{-1}.
\]
This gives
\[
  \bmxl{} = -2 \tril\left( \mxb{}\sym(\bmxb{})\mxl{}^{-T} \right).
\]
This can be done in-place. Here, outputs cannot overwrite inputs. In fact, our implementation uses
\[
  2\mxb{}\sym(\bmxb{}) = \mxb{}\bmxb{} + \mxb{}\bmxb{}{}^T,
\]
even if this is slightly wasteful, in order to avoid having to allocate temporary memory for $\sym(\bmxb{})$.

If the triangular matrix $\mxr{}$ is upper triangular instead, and $\mxa{} = \mxr{}^T\mxr{}$, we plug $\mxl{}=\mxr{}^T$ into the derivation, to obtain
\[
  \bmxr{} = -2\triu\left( \mxr{}^{-T}\sym(\bmxb{})\mxb{} \right).
\]

\end{document}